\documentclass[prb,preprint]{revtex4-1} 

\usepackage{amsmath}  
\usepackage{amsfonts} 
\usepackage{graphicx} 
\usepackage{color}
\usepackage{enumitem}

\newcommand{\dt}{\Delta t}
\newcommand{\dx}{\Delta x}
\newcommand{\dxdet}{\Delta x_\mathrm{det}}
\newcommand{\dxrand}{\Delta x_\mathrm{rand}}

\newcommand{\frand}{f_\mathrm{rand}}

\newcommand{\kt}{k_B T}
\newcommand{\lpar}{\left(}
\newcommand{\mfpt}{\mathrm{MFPT}}
\newcommand{\msd}{\mathrm{MSD}}

\newcommand{\rpar}{\right)}
\newcommand{\tij}{T_{i \! \to \! j}}
\newcommand{\tmax}{t_\mathrm{max}}
\newcommand{\traj}{\mathrm{traj}}
\newcommand{\tss}{t_\mathrm{SS}}
\newcommand{\xtraj}{\left\{ x_0, x_1, x_2, \ldots \right\}}
\newcommand{\xtrajn}{\left\{ x_0, x_1, x_2, \ldots , x_N \right\}}
\newcommand{\xvec}{\vec{x}}

\renewcommand{\eqref}[1]{Eq.~(\ref{#1})}

\begin{document}


\title{A gentle introduction to the non-equilibrium physics of trajectories: Theory, algorithms, and biomolecular applications}

\author{Daniel M. Zuckerman}
\email{zuckermd@ohsu.edu} 
\author{John D. Russo}
\affiliation{Department of Biomedical Engineering, Oregon Health \& Science University, Portland, OR 97239}


\date{\today}

\begin{abstract}
Despite the importance of non-equilibrium statistical mechanics in modern physics and related fields, the topic is often omitted from undergraduate and core-graduate curricula.  Key aspects of non-equilibrium physics, however, can be understood with a minimum of formalism based on a rigorous trajectory picture.
The fundamental object is the ensemble of trajectories, a set of independent time-evolving systems, which easily can be visualized or simulated (e.g., for protein folding) and which can be analyzed rigorously in analogy to an ensemble of static system configurations.  
The trajectory picture provides a straightforward basis for understanding first-passage times, “mechanisms” in complex systems, and fundamental constraints on the apparent reversibility of complex processes.  
Trajectories make concrete the physics underlying the diffusion and Fokker-Planck partial differential equations.  
Last but not least, trajectory ensembles underpin some of the most important algorithms that have provided significant advances in biomolecular studies of protein conformational and binding processes.

\end{abstract}

\maketitle 


\section{Introduction} 

Most of the phenomena we encounter in daily life, from weather to cooking to biology, are fundamentally out of equilibrium and require physics typically not touched on in the undergraduate or even graduate physics curricula.
Many physics students are alarmed at the complexity and abstraction of thermodynamics and ``sadistical mechanics,'' and understandably would not seek out instruction in non-equilibrium statistical physics.
Yet there is a surprising range of fundamental non-equilibrium material that can be made accessible in a straightforward way using \emph{trajectories}, which are essentially movies of systems executing their natural dynamics.
The trajectory picture first and foremost is fundamental \cite{onsager1931reciprocal,vanden2010transition,elber2020molecular} --- for example, dynamics generate equilibrium, but not the other way around. \cite{zuckerman2010statistical}
It can also lead, with a minimum of mathematics, to understanding key non-equilibrium phenomena (relaxation and steady states) and similarly to extremely powerful cutting-edge simulation methods (path sampling).
Students deserve a taste of this material.

Why are trajectories fundamental?
A trajectory is simply the sequence of phase-space points through which a system passes, recorded perhaps as a ``movie'' listing all atomic positions and velocities at evenly spaced time points -- the ``frames'' of the movie.
Such movies are fundamental because, as we learned from Newton, nature creates forces that lead to dynamics, \cite{giancoli2008physics} i.e., to trajectories.
We may attempt to describe the dynamics in various average ways -- e.g., using equilibrium ideas -- but the trajectories are the basis of everything.
Theories, such as equilibrium statistical mechanics, generally build in assumptions, if not approximations.
In fact, the most fundamental definition of equilibrium itself derives from dynamics, via detailed balance, \cite{onsager1931reciprocal,reif2009fundamentals,zuckerman2010statistical} whereby there must be an equal-and-opposite balance of flows between any two microstates.

Dynamical descriptions generally have more information in them than average or equilibrium theories. \cite{seifert2012stochastic,zuckerman2010statistical,Zuckerman2020keybio}
As a simple example, perhaps you know that someone sleeps eight hours a day.
But that average hides the time at which sleep occurs, as well as whether it includes an afternoon nap.
In the case of diffusion, we know that particles observed in a localized region will tend to spread out over time.
But if we only observe the spatial density, we don't know which particles went where.
Trajectories, which track particles over time, inherently capture this information.

A trajectory ensemble description, as described below, provides \emph{the} key observables for transition processes: rate and mechanism.
In a biomolecular context, these are essentially everything we want to know.
Consider protein folding.
We want to know how fast proteins fold and how folding rates change under specific mutations. \cite{lindorff2011fast,fersht1999structure}
We also want to know the mechanism of folding: the conformations that are visited during the process which in turn can illuminate chemical-structure causes of rate changes due to mutation. \cite{juraszek2006sampling,fersht1999structure}
Other conformational processes in biomolecules arguably are of even greater interest, such as binding \cite{mobley2017predicting} and allostery \cite{gunasekaran2004allostery,fersht1999structure}, due to their implications for drug design; here again rate and mechanism are of utmost importance. \cite{zwier2016efficient,copeland2006drug}

This article will explain the theory of trajectory ensembles, starting with simple diffusion and moving to systems with complex energy landscapes.
We will explore essential aspects of non-equilibrium statistical mechanics, focusing on timescale quantification via the mean first-passage time.
The understanding of non-equilibrium trajectory ensembles leads directly to the ``super parallel'' weighted ensemble simulation methodology, widely used in computational biology \cite{zuckerman2017weighted}, which is explored in a one-dimensional pedagogical example.
A number of exercises are given along with clearly demarcated more advanced material.

The statistical mechanics of trajectories has been addressed pedagogically, in different ways, in prior work.
Clear, basic-level descriptions can be found in some textbooks \cite{zuckerman2010statistical,elber2020molecular} and path-sampling papers in the molecular-oriented literature. \cite{pratt1986statistical,dellago1998transition,zuckerman1999dynamic}
Astumian and coworkers highlighted the importance of trajectories and their probabilistic description in multiple contexts, \cite{bier1999intrawell,astumian2006unreasonable} and provided important semi-microscopic, discrete-state descriptions of molecular motors, \cite{astumian2010thermodynamics,astumian2020nonequilibrium} building on the seminal work of Hill. \cite{hill1982linear,hill2004free}
Phillips and coworkers employed trajectory concepts in presenting Jaynes's maximum-caliber approach to inferring kinetics; \cite{ghosh2006teaching} note related work by Dill and co-workers. \cite{presse2013principles,ghosh2020maximum}
Swendsen's discussion of irreversibility is also of interest, \cite{swendsen2008explaining} as is the classic treatment by Chadrasekhar. \cite{chandrasekhar1943stochastic}
The present discussion attempts to provide a more elementary discussion of trajectory physics, with a focus on computational applications not found in most prior work.
Perhaps unexpectedly, the path sampling algorithms derivable from the present description are very much at the leading edge of molecular computation. \cite{chong2017path}

\section{Basics: Dynamics and trajectories}
\label{sec:diffusion}
In this section, we introduce the building blocks of our analysis, starting from one-dimensional Newtonian motion.
We add fundamental stochastic elements, and then develop the trajectory picture with an associated numerical recipe.

\subsection{Stochastic dynamics}

The starting point for our quantitative trajectory description is the simplest form of stochastic dynamics, often called Brownian dynamics, which we will justify starting from Newton's second law.
Brownian dynamics are also known by more intimidating terminology, as overdamped Langevin dynamics, but their essence is simple to understand.
As a familiar reference, we first write the one-dimensional (1D) law of classical motion,
\begin{equation}
    m \frac{dx^2}{dt^2} = f \;,
    \label{newton}
\end{equation}
where $m$ is mass, $x$ is position, and $f = - dU/dx$ is the force, with $U(x)$ the potential energy.
Advancing one step in complexity, the 1D Langevin equation models motion in a viscous (frictional) medium by adding a damping force that always opposes the direction of motion (velocity), as well as a random force $\frand$ from collisions, \cite{reif2009fundamentals,zuckerman2010statistical} yielding
\begin{equation}
    m \frac{d^2x}{dt^2} = f - \gamma \, m \frac{dx}{dt} + \frand \;,
    \label{langevin}
\end{equation}
where $\gamma > 0$ is the friction constant -- effectively, a collision frequency, as can be seen by dimensional analysis.
Details of the random force will be given later.
Both forces are needed, otherwise damping would eliminate all motion.

In the \emph{overdamped} limit, inertia is ignored.  This is akin to motion in a beaker filled with thick oil: there is minimal tendency for an object to continue in any given direction in the absence of force; with a force such as gravity, terminal (constant) velocity is reached quickly -- i.e., no further acceleration occurs despite the force.  
At microscopic scales, however, there continues to be random thermal motion due to molecular collisions.
Setting the inertial term $m \,d^2x/dt^2$ to zero in \eqref{langevin} and re-arranging terms, the overdamped Langevin equation is \cite{reif2009fundamentals,zuckerman2010statistical}
\begin{equation}
    \frac{dx}{dt} = \frac{1}{m\gamma} \lpar f + \frand \rpar \;.
    \label{overdamped}
\end{equation}
This simplified equation of motion may look unusual to those unfamiliar with it, but studying its application in a numerical context will make its physical basis and relation to diffusion more clear.

\subsection{Time-discretized overdamped dynamics and computation}

\begin{figure}[h]
    \centering
    \includegraphics[width=10cm]{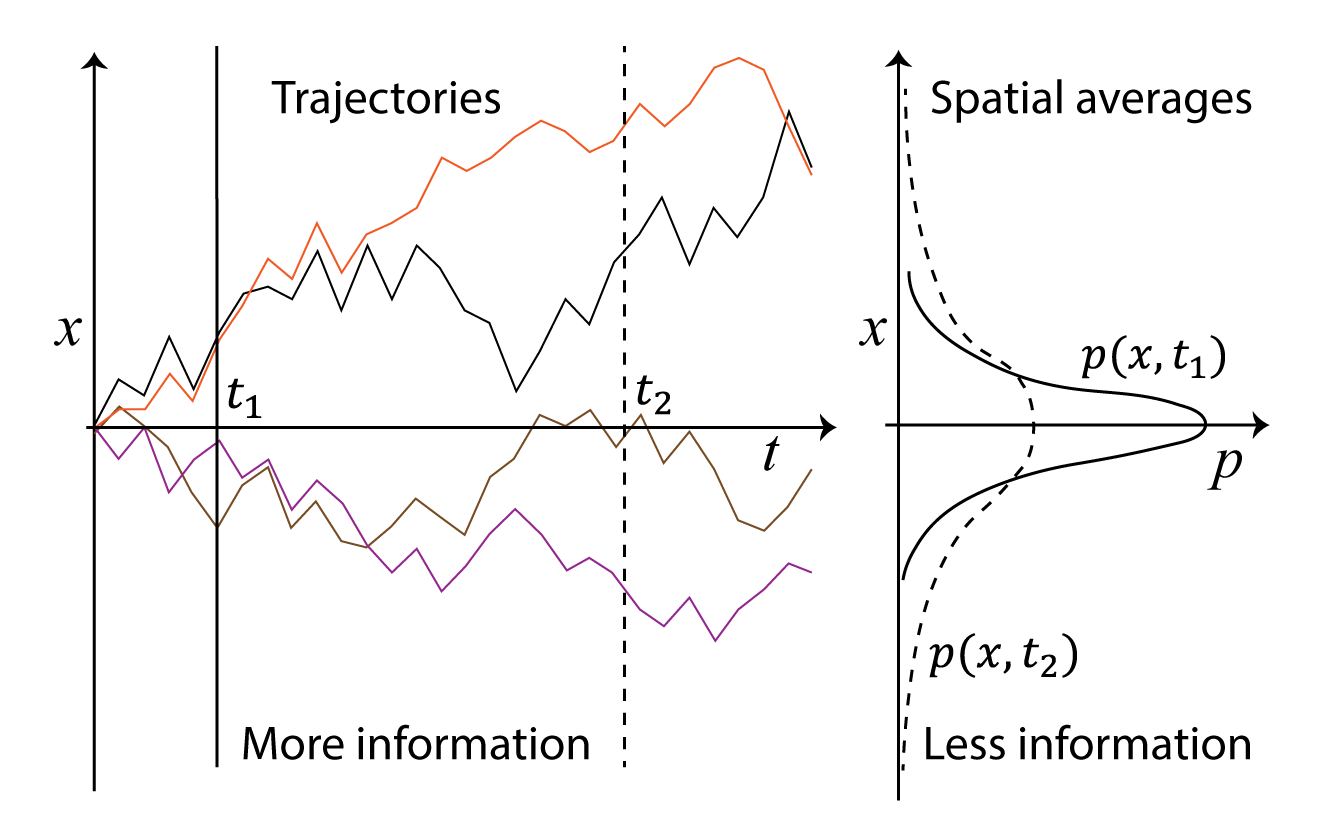}
    \caption{Simple diffusion, two ways.  At left are schematic time-discretized trajectories illustrating one-dimensional diffusion started from the initial point $x_0 = 0$.  Averaging over the positions of many trajectories at specific time points $t_1$ and $t_2$ yields the distributions shown at right, with $p(x, t_i) = p(x_i | x_0)$.
    Averaging can aid interpretation but it also removes information, namely, the  connectivity among the trajectories' sequences of points.}
    \label{fig:diff_traj}
\end{figure}

We will make most use of a discrete-time picture (fixed time steps) which not only greatly simplifies the mathematics but also translates directly into simple computer implementation.
If we discretize the dynamics of \eqref{overdamped} by writing the velocity as $\dx/\dt$ and multiplying through by $\dt$, we arrive at a very useful equation,
\begin{equation}
    \dx = \frac{\dt}{m\gamma} \lpar f + \frand \rpar = \dxdet + \dxrand \;,
    \label{dxoverd}
\end{equation}
where 
    $\dxdet = f \dt/m\gamma$
is the deterministic component of the spatial step due to an external force (e.g., molecular, gravitational or electrostatic) and $\dxrand$ is the random part due to thermal molecular collisions.
At finite temperature, microscopic motion must not cease and hence, in the Langevin picture, thermal fluctuations must balance ``dissipation'' due to damping of the $\gamma$ term. \cite{vanKampen1992stochastic,reif2009fundamentals}  To accomplish this, $\dxrand$ is typically assumed to follow a zero-mean Gaussian distribution which must have its variance given by \cite{zuckerman2010statistical}
\begin{equation}
    \sigma^2 = 2 \kt \dt / m \gamma \; .
    \label{sig}
\end{equation}
The Gaussian assumption is justified on the basis of the central limit theorem, \cite{zuckerman2010statistical} because a molecule in aqueous solution can experience upwards of $10^{13}$ collisions per second, \cite{reif2009fundamentals,chandrasekhar1943stochastic} and hence a large number of collisions occur in any $\dt > 1$ ns.
The high collision frequency also justifies the implicit assumption here that sequential $\dxrand$ values are independent, i.e., not time-correlated.

With the distribution of $\dxrand$ specified, the discrete overdamped dynamics \eqref{dxoverd} is simultaneously a prescription for computer simulation of trajectories \emph{and} directly implies a probabilistic description of trajectories.
Let's start with computer simulation, which is simpler by far.
Defining $x_j = x(t\!=\!j\dt)$,
Eq.\ \eqref{dxoverd} is essentially a recipe for calculating the next position $x_{j+1} \equiv x_j + \dx$ in a time-sequence, given $x_j$.
For a sufficiently small time step $\dt$, the force $f$ will be approximately constant over the whole time interval, so we take
\begin{equation}
    \dxdet = f(x_j) \dt/m\gamma 
    \label{dxdet}
\end{equation}
and $\dxrand$ is chosen from a Gaussian (normal) distribution of variance $\sigma^2$ from \eqref{sig}.
Looping over this process yields a discrete-time \emph{trajectory}:
\begin{equation}
    \traj = \xtraj \; ,
    \label{trajone}
\end{equation}
which is just a list of positions at intervals of $\dt$.
We can easily recast trajectory elements in terms of spatial increments,
\begin{align}
    x_0 &= x_0 \hspace{0.5cm} \mbox{(arbitrary)} \nonumber \\
    x_1 &= x_0 + \dx_1  \label{dxtraj} \\ 
    x_2 &= x_0 + \dx_1 + \dx_2 = x_1 + \dx_2 \nonumber \\
    \cdots \; ,\nonumber
\end{align}
which is useful for understanding simulation algorithms such as \eqref{dxoverd}.

Trajectories of \emph{simple diffusion} can be generated from Eq.\ \eqref{dxoverd} by setting $f=0$ (hence $\dxdet=0$).
The recipe given above simplifies to choosing a Gaussian random step at each time point, i.e.,
\begin{equation}
    \dx = \dxrand  \hspace{1cm} \mbox{(simple diffusion)}
    \label{dxdiff}
\end{equation}
as we would expect.
Schematic examples of these simplest stochastic trajectories are shown in Fig.\ \ref{fig:diff_traj}.
There is no directionality in simple diffusion, but only a statistical tendency to diffuse away from the starting point, as will be quantified below.


\section{Simple diffusion in the trajectory picture}

The basics of diffusion, such as Fick's law and the diffusion equation, are well known, so diffusion theory is a perfect context for introducing the trajectory formulation.  
Students may find that following the behavior of individual particles is a more concrete exercise than visualizing probability distributions.
In this section, we show that the trajectory approach yields the familiar average description of simple diffusion in a force-free (constant-energy) landscape.
In the bigger picture, we get an explicit sense of physical details of trajectories which are averaged (integrated) out to yield the distribution picture.

\subsection{Probabilistic picture for trajectories}
We start by analyzing diffusive trajectories based on random steps where the force $f$ has been set to zero.
The procedure \eqref{dxdiff} of repeatedly choosing a Gaussian step with variance from \eqref{sig} implicitly but precisely defines a probability distribution for an entire trajectory \eqref{trajone}, which will prove of fundamental importance.
First, by construction, the probability of a \emph{single} step $\dx$ is given by
\begin{equation}
    p_1(\dx) = \frac{1}{\sigma\sqrt{2\pi}} e^{-\dx^2 / 2 \sigma^2} \; .
    \label{ponediff}
\end{equation}
This is the meaning of choosing a Gaussian step.
Note that \eqref{ponediff} depends only on the magnitude and not on the starting point of the specific step, which is a characteristic of simple diffusion because no forces are present.

For the full trajectory, we use the simple rule that the probability of a sequence of independent steps is simply the product of the individual step probabilities: think of a sequence of fair coin flips characterized by $1/2$ to the appropriate power.
Hence, for an $N$-step trajectory defined by \eqref{dxtraj} starting from $x_0$, we have
\begin{align}
        p(\traj) &= p_1(\dx_1) \cdot p_1(\dx_2) \, \cdots \, p_1(\dx_N)
        \label{trajdistone} \\
        &= \lpar \frac{1}{\sigma\sqrt{2\pi}} \rpar^{\!\!N} \prod_{j=1}^N e^{-\dx_j^2 / 2 \sigma^2}
        \label{trajdist}
\end{align}

A multi-dimensional distribution such as \eqref{trajdist} may not be trivial to understand for those not used to thinking in high dimensions.
First, why is it a multi-dimensional distribution?
Well, it describes the distribution of a \emph{set} of points, the trajectory $\xtrajn$.
Note that we immediately obtain the $\dx$ values needed for \eqref{trajdist} from the $x$ values using \eqref{dxtraj}: $\dx_1 = x_1 - x_0$ and so on.
So if you're given a set of (trajectory) $x$ values, you can convert them into $\dx$ values and plug them into \eqref{trajdist} to get the probability of that trajectory.
You can do this for \emph{any} set of $x$ values, even ridiculously unphysical values with gigantic jumps -- but of course the probability will be tiny for unphysical trajectories.
For completeness, strictly speaking, \eqref{trajdist} is a probability \emph{density} \cite{zuckerman2010statistical} and absolute probabilities are only obtained by integrating over a finite region.

The distribution of trajectories encodes all the information we could possibly want about diffusive behavior, although some math is needed to get it.
Alternatively, as a proxy for the distribution, multiple trajectories could be simulated to quantify their average behavior.
In the case of simple diffusion, however, the math of the trajectory distribution is both tractable and illuminating.

As a fascinating technical aside, note that the product of exponentials in \eqref{trajdist} can be re-written as the exponential of a sum ($-\sum_j \dx_j^2/2\sigma^2$), which makes the probability look somewhat like a Boltzmann factor.
Indeed, consulting the definition of $\sigma^2$ in \eqref{sig} we find it is proportional to $k_B T$.  Of course, the argument of our exponential is not a true energy, but can be considered an effective path energy, known as the ``action.'' \cite{onsager1953fluctuations,zuckerman2010statistical}
(In the non-diffusive case, $\dxdet \neq 0$ leads to an additional term in the exponent and the action; see below.)
The action formulation, and the consideration of all possible paths, is the heart of the path-integral formulation of quantum mechanics. \cite{Sakurai1985modern}
The path-probability formulation is truly fundamental to physics.

\subsection{Deriving the spatial distribution from trajectories}

A key observable of interest is the distribution of $x$ values at a fixed but arbitrary time point (Fig.\ \ref{fig:diff_traj}).
To build up to this, we'll carefully derive the equation for the conditional probability distribution $p(x_2 | x_0)$ of $x_2 = x(2\dt)$ values -- i.e., the distribution for a fixed starting point $x_0$.
The critical idea is that we can obtain the probability of any given $x_2$ value by summing (i.e., integrating) over all possible two-step trajectories that reach the particular value starting from $x_0$.
Because both forward and backward motion are possible for the intermediate step, we must consider all possible $x_1$ values.
Mathematically, this amounts to
\begin{align}
    p(x_2 | x_0 ) &= \int_{-\infty}^{\infty} dx_1 \, p_1(\dx_1) \, p_1(\dx_2) \nonumber \\
    &= \int_{-\infty}^{\infty} d\dx_1 \, p_1(\dx_1) \cdot p_1(x_2 - (x_0+\dx_1)) \, , 
    \label{ptwo}
\end{align}
where we have used \eqref{trajdistone} to start and then \eqref{dxtraj} to substitute for $\dx_2$.

We can evaluate the integral in \eqref{ptwo} exactly.
Plugging in the expression for $p_1$ from \eqref{ponediff} and setting $y=\dx_1$, we have
\begin{align}
    p(x_2| x_0) &= \frac{1}{2\pi \sigma^2} \int_{-\infty}^{\infty} dy \, e^{-y^2 / 2 \sigma^2} e^{-(x_2 - x_0 - y)^2 / 2 \sigma^2} \nonumber \\
    &= \frac{1}{2\pi \sigma^2} \,
    e^{-(x_2-x_0)^2/4 \sigma^2} \!
    \int_{-\infty}^{\infty} dy \, e^{-[y-(x_2-x_0)/2]^2/\sigma^2} \nonumber \\
    &= \frac{1}{\sqrt{2\pi} \lpar \sqrt{2} \sigma \rpar} \,
    e^{-(x_2-x_0)^2/2 \cdot 2 \sigma^2} \; ,
    \label{ptwoposn}
\end{align}
where the second line is derived by completing the square in the exponent and the third line is derived by performing the Gaussian integral shown.

The result \eqref{ptwoposn} for the distribution of positions after $2\dt$ is very informative, especially by comparison to the single-step distribution \eqref{ponediff}.
The distribution of possible outcomes is still a Gaussian of mean $x_0$, but the variance is doubled -- equivalently, the standard deviation has increased by a factor of $\sqrt{2}$.
See Fig.\ \ref{fig:diff_traj}.
It is important that \emph{we derived this distribution by averaging (integrating) over the ensemble of two-step trajectories}.
As promised, the information was indeed encoded in the original trajectory distribution \eqref{trajdist}.

From here, it's not hard to generalize to an arbitrary number of steps by repeating the integration process.
The result is that the distribution of $x_n = n\dt$ values is also a Gaussian with mean $x_0$, but with variance $n \sigma^2$:
\begin{equation}
    p(x_n| x_0) = \frac{1}{\sqrt{2n\pi} \sigma} \,
    e^{-(x_n-x_0)^2/ 2 n\sigma^2} \; ,
    \label{pnposn}
\end{equation}
Eq.\ \eqref{pnposn} embodies the usual description of diffusion, as we will see in two ways, but it also contains \emph{less} information than our initial trajectory description.

\subsection{Confirming the probabilistic description of diffusion}
Have we really recapitulated the usual description of diffusion?
As a first check, we immediately recover the expected linear time dependence of the mean-squared displacement \cite{zuckerman2010statistical} based on \eqref{pnposn}.
This is because the variance \emph{is} the mean-squared displacement or deviation ($\msd$) and the number of steps $n=t/\dt$ is simply proportional to time.
By the definition of a Gaussian distribution, the variance implicit in \eqref{pnposn} is $n \sigma^2$ and we therefore have
\begin{align}
    \msd & \equiv \left \langle \lpar x_n - x_0 \rpar^2 \right \rangle = \int_{-\infty}^\infty dx_n \, \lpar x_n - x_0 \rpar^2 \, p(x_n|x_0)
     \nonumber \\
    &= n \sigma^2 = (t/\dt) \sigma^2 \; .
    \label{dconst}
\end{align}
If we define the diffusion constant via $\msd = 2D \, t$ (in one dimension), then from \eqref{sig} and \eqref{dconst},  we derive $D = \kt / m \gamma$, which is a well-known result. \cite{zuckerman2010statistical}

Second, by renaming the variable $x_n \to x = x(t)$ in \eqref{pnposn} and noting that time $t = n\dt$, we can see that \eqref{pnposn} describes the time-evolving probability distribution of positions $p(x,t)$, which is the well-known solution to the 1D diffusion equation,
\begin{equation}
    \frac{\partial p}{ \partial t} = D \frac{ \partial^2 p}{\partial x^2} \; ,
    \label{diffpde}
\end{equation}
This can be verified by direct differentiation, but see Exercises below for a hint.
The agreement with the \emph{continuous}-time diffusion equation implies that time discretization is irrelevant, but be warned that this is not always the case, as discussed in the Exercises.


\subsection{What is missing from the standard description of diffusion?}
Because the distribution of positions \eqref{pnposn} is known for any time and provides the exact solution to the diffusion equation, it may seem there is nothing more to know.
But the key observables --- the timescale (or rate) and mechanism of any particular process --- either are not available at all from the positional distribution, or not easily available. \cite{gardiner2009stochastic,Risken1996}

These shortcomings stem from the information missing from the spatial distribution.
Even if we know the spatial distribution at two times, \emph{we still do not know how any given diffusing particle went from one place to another.}
That is, although we know the fraction of particles that will be located between any $x$ and $x+dx$, we do not know which came from left or right and exactly from where.
This information is encoded in the dynamics, and recorded in the distribution of trajectories \eqref{trajdistone}, which is essentially a distribution of paths taken through position space.
It is fair to say, therefore, that the trajectory distribution \emph{is} the mechanism, assuming that all trajectories considered conform to criteria of interest (e.g., starting at $x=0$ and perhaps reaching a value $x>a$ after $n$ steps.)

\subsection{Beyond simple diffusion in one dimension}

Before we move beyond a single dimension, a useful reference for developing intuition is the generalization of the single-step distribution \eqref{ponediff} when a force is present.
Re-framing the procedure \eqref{dxoverd} probabilistically, the distribution for overdamped dynamics of a 1D particle in the presence of a spatially varying potential $U(x)$ is a different Gaussian:
\begin{equation}
    p_1(\dx) = \frac{1}{\sqrt{2\pi}\sigma} e^{-\lpar \dx - \dxdet \rpar^2 / 2 \sigma^2} \; .
    \label{pone}
\end{equation}
In contrast to the simple-diffusion case \eqref{ponediff}, the distribution of possibilities is centered on the deterministic (force-driven or ``drift'') step $\dxdet$ defined by \eqref{dxdet}.
That is, the particle tends to move in the direction of the force, albeit stochastically.

Eq.\ \eqref{pone} should guide your intuition for single-step motion of a stochastic system: there is a distribution of possibilities centered on the deterministic step.
The deterministic component generally could depend on inertia and/or force, although in the overdamped case there is no inertia.
Note that $\dxdet$ in \eqref{pone} implicitly depends on the starting position for the step: see \eqref{dxdet}.
Below, we will make the position dependence more explicit.



\section{The non-equilibrium steady state (NESS) and the Hill relation for rates}
\label{sec:ness}

Probably the most important observable in a dynamical process, at least in biomolecular studies, is the rate for a process.
As we will see, the rate is closely related to a specific non-equilibrium steady state, which is essential to understand but also quite accessible.

\begin{figure}[h]
    \centering
    \includegraphics[width=8cm]{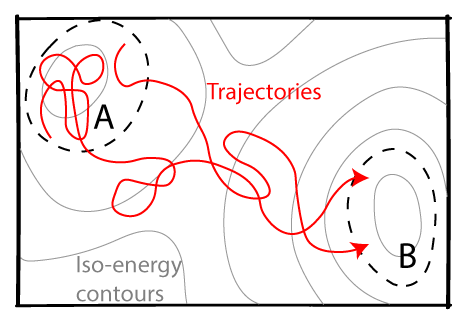}
    \caption{Source-sink non-equilibrium steady state.  Trajectories (red curves, color online) are initiated in state A and terminated upon reaching state B, with states bounded by dashed contours.  Importantly, trajectories that reach B are then re-initiated from A.  Such a system will reach a non-equilibrium steady state after a transient ``relaxation'' period.  Gray solid lines show iso-energy contours of a schematic landscape.}
    \label{fig:ness}
\end{figure}

Physicists often quantify a rate via the mean first-passage time (MFPT).  \cite{redner2001guide,vanKampen1992stochastic,Risken1996,gardiner2009stochastic}
The first-passage time is simply the time required for a process from start to finish -- e.g., the time required for a protein to fold, starting from when it is initialized in an unfolded state.
In Fig.\ \ref{fig:ness}, this is the time from initiation in ``source'' state A to absorption in ``sink'' state B.
(We are thus employing source-sink boundary conditions.)
Chemists and biochemists quantify kinetics via the ``rate constant'' for a conformational process like protein folding, which has units of s$^{-1}$ and can be defined as the reciprocal MFPT, although chemists prefer a definition based on directly measurable ``relaxation times.'' \cite{zuckerman2010statistical,chandler1987introduction,chodera2011robust}
Our discussion will focus solely on the MFPT for simplicity.

The MFPT can be directly obtained from a steady-state trajectory ensemble, so we will start by defining a source-sink non-equilibrium steady state (NESS) as sketched in Fig.\ \ref{fig:ness}.
Independent trajectories are initiated in the source macrostate A (e.g., the set of unfolded protein configurations) according to a specified distribution $p_0$ (e.g., a single configuration or the equilibrium distribution over A).
A second, non-overlapping sink macrostate B is an absorbing state in that trajectories reaching B are terminated, although in our source-sink setup they are immediately restarted in A selected according to the $p_0$ distribution.
If this process is allowed to run for long enough so that each trajectory has reached B and been recycled back to A many times, the system will reach a non-equilibrium steady state.
Without a sink state or recycling, the system will relax to equilibrium, which is also a steady state.
(See Exercises below to explore the difference between equilibrium and other steady states.)

The MPFT is derivable from a NESS trajectory ensemble in a direct way which will seem obvious once you're aware of it.
The derivation is simple, but requires some thought.
Imagine we have a large number $M \gg 1$ of independent systems that together make up the source-sink NESS (Fig.\ \ref{fig:ness}).
By construction, the NESS is characterized by a \emph{constant flow} of trajectories into B.
We can simply count the number of trajectories arriving during some time interval $\tau$, and we'll call this count $m$.
Thus, a fraction $m/M$ of the total probability arrives in time $\tau$.

To continue our derivation, we can estimate this same fraction of trajectories arriving based solely on the meaning of the MFPT.
By definition, the average amount of time a trajectory requires to traverse from A to B \emph{is} the MFPT, so the (average) probability for any given trajectory to arrive during an interval $\tau$ is precisely $\tau/\mfpt$ ... which in turn is the same as the fraction expected to arrive in $\tau$.
In other words, $m/M = \tau/\mfpt$ and we have therefore derived the \emph{Hill relation} \cite{hill2004free,zuckerman2010statistical}
\begin{equation}
    \frac{1}{\mfpt} = \frac{m/M}{\tau} = \mbox{Flux}(A \to B \, | \mbox{NESS}) \; ,
    \label{hill}
\end{equation}
where the flux is the probability arriving to B per unit time in the NESS.
Eq.\ \eqref{hill} is an exact relation with no hidden assumptions, although not surprisingly the MFPT is particular to the initiating distribution $p_0$ of the particular NESS in which the flux is measured.
That is, the MFPT depends on where in state A trajectories are initiated.

The Hill relation hints at a remarkable possibility: estimation of a long timescale (the MFPT) based on an arbitrarily short period of observation ($\tau$).
If this could be done routinely, it would represent a major accomplishment in computational physics. \cite{chong2017path}
Below in Sec. \ref{sec:we}, we describe a simple algorithm that can leverage the Hill relation for practical computations in many systems.
We also explain the challenges involved.

\section{More advanced discussion of ensembles and thermodynamic states}

This section describes additional fundamental concepts in non-equilibrium physics, but the discussion necessarily becomes more technical.
Readers can skip this section without compromising their ability to understand subsequent material.

\subsection{Notation and nomenclature for multi-dimensional systems}


We will frame our discussion a bit more generally, in the context of multi-dimensional systems.  Fortunately, this extension adds only incremental conceptual and mathematical complexity.  To keep notation as simple as possible, we will use $\xvec$ to represent \emph{all} microscopic coordinates --- the phase-space vector consisting of all positions and velocities of all atoms in our classical representation.
In some cases, such as overdamped dynamics \eqref{overdamped}, velocities may be excluded from the description but the $\xvec$ notation remains valid.
%
A \emph{macrostate} 
is defined to be a \emph{set} of $\xvec$ points.
These macrostates are not to be confused with thermodynamic states such as equilibrium at some constant temperature or a non-equilibrium steady state.

As with our discussion of simple diffusion above, we will strictly use discrete time: $t=0, \dt, 2\dt, ...$.
Discrete time greatly simplifies our description of trajectory probabilities without sacrificing any physical insights.
In a trivial extension of \eqref{trajone}, we therefore write a trajectory as 
\begin{equation}
    \traj = \left\{ \xvec_0, \, \xvec_1, \, \xvec_2, \ldots \right\} \; ,
    \label{traj}
\end{equation}
where $\xvec_j$ is the phase point at time $t=j\dt$.

\subsection{The initialized trajectory ensemble in multiple dimensions}


The probabilistic description of a multi-dimensional trajectory follows logic almost identical to the 1D diffusion formulation \eqref{trajdistone}, except for two details.
First, we now include the possibility that the initial system phase point $\xvec_0$ itself is chosen from some distribution $p_0$.
Second, in contrast to simple diffusion, where the distribution \eqref{ponediff} of outcomes $p_1$ for any single step depends only on the magnitude of $\dx$, more generally the outcome depends on the starting point of the step because the force may vary in space.
We therefore adopt a notation which makes this explicit: $p_1(\xvec_{j-1} \to \xvec_j) = p_1(\xvec_j | \xvec_{j-1})$ is the (conditional) probability distribution for $\xvec_j$ values, given the prior position $\xvec_{j-1}$.
The probability of a full trajectory is then the product of the initial distribution and the sequence of stepwise distributions,
\begin{equation}
    p(\traj) = p_0(\xvec_0) \cdot p_1(\xvec_0 \to \xvec_1) \cdot p_1(\xvec_1 \to \xvec_2) \, \cdots \, p_1(\xvec_{N-1} \to \xvec_N) \; .
    \label{trajdistvec}
\end{equation}
The mathematical form of $p_1$ must now account for multi-dimensional aspects of the system, as well as any forces or inertia if present: see \eqref{pone} and the discussion following it.
Although specifying $p_1$ in generality is beyond the scope of our discussion, we should note that the form \eqref{trajdistvec} indicates we have assumed Markovian behavior: the distribution of outcomes $p_1$ at any time depends only on the immediately preceding time point.

We must be careful to specify our system without ambiguity.
A given physical system, such as a particular protein molecule in a specified solvent at known temperature and pressure, can be considered in a variety of thermodynamic states, such as equilibrium or a non-equilibrium state.
The system and the thermodynamic conditions both must be specified.
Conveniently, the two aspects are described by different parts of the trajectory distribution equation \eqref{trajdistvec}:
the intrinsic physical properties such as forces and dynamics are encoded in the single-step $p_1$ factors, while the thermodynamic state or ensemble is determined by the initial distribution $p_0$ along with boundary conditions.
Some boundary conditions will be discussed below.

The distribution \eqref{trajdistvec} describes the \emph{initialized trajectory ensemble}, the set of trajectories originating from a specified phase-point distribution $p_0$ at time $t=0$.
For instance, $p_0$ could represent a single unfolded protein configuration (making $p_0$ a Dirac delta function), a set of unfolded configurations, a solid in a metastable state, or the set of initial positions of multiple dye molecules in a solvent.
Fig.\ \ref{fig:diff_traj} illustrates the one-dimensional trajectory ensemble initialized from $p_0(x) = \delta(x)$.

As with simple diffusion, we can revert to the simpler, averaged description of a spatial distribution that evolves in time due to the dynamics.
That is, in principle we can calculate the distribution of phase points at time $t=N\dt$ starting from $p_0$, denoted $p(\xvec_N | \, p_0)$.
When forces are present, the diffusion (partial differential) equation \eqref{diffpde} must be generalized to account for the tendency for a particle to move a certain direction, leading to the Fokker-Planck/Smoluchowski picture \cite{gardiner2009stochastic,Risken1996,vanKampen1992stochastic}.
Appendix \ref{app:smol} describes the corresponding Smoluchowski equation that governs overdamped motion with forces.
However, as with simple diffusion, the spatial distribution represents an average over the information-richer trajectories.

\subsection{Connection to relaxation, state populations, and thermodynamics}

It is important to note that, in general, an initialized system will ``relax'' away from its initial distribution $p_0$.
For systems of interest here, the system's phase-point distribution $p(\xvec_N | \, p_0)$ will tend to relax toward a \emph{steady state} dependent only on the boundary conditions.
In a constant-temperature system with no particle exchange, for example, the distribution will approach equilibrium as embodied in the Boltzmann factor:
$\lim_{N\to\infty} p(\xvec_N | \, p_0) \propto \exp(- H(\xvec) / k_B T)$, where $H(\xvec)$ is the total energy of point $\xvec$, $k_B$ is Boltzmann's constant and $T$ is the absolute temperature.
In general, whether equilibrium or not, the steady state that is reached typically will be independent of $p_0$ after sufficient time for a `well-behaved' system.
In Sec.~\ref{sec:ness}, we explored \emph{non-equilibrium} steady states critical to understanding conformational transitions.

Whether the system is in the relaxation or steady regime, the phase-point distribution $p(\xvec)$, obtainable from the trajectory picture, directly connects to observable and thermodynamic properties.
Most simply, the time-dependent macrostate population can be obtained as the integral of $p(\xvec)$ over a region of phase space: this is the fraction of probability in the state which evolves in time with $p$.
At a system-wide level, both the entropy and average energy can be obtained from well-known integrals over $p$. \cite{seifert2012stochastic,zuckerman2010statistical}
These also evolve with time, directly leading to the entropy production picture.
Further detail on these topics is beyond the scope of the present discussion, and interested readers should consult suitable references.  \cite{seifert2012stochastic,presse2013principles}


\subsection{The ensemble of trajectories and the meaning of equilibrium}
\label{sec:equil}

When we speak of an ``ensemble'' of trajectories, the word has the same meaning as in ordinary statistical mechanics, \cite{reif2009fundamentals,zuckerman2010statistical}namely, a set of \emph{fully independent} trajectories generated under the conditions of interest (see below).
That is, each member of the ensemble is a replica of the same physical system but is initiated from a phase point that typically will differ from others in the ensemble.

An ensemble in principle can be generated according to any process and under any conditions we care to specify.
The dynamics of these trajectories could be governed by simple constant-temperature diffusion, or there could be a temperature gradient, forces, or both.
Trajectories could additionally be subject to certain boundary conditions: for example, they might be assumed to reflect off some boundary in phase space, or be absorbed on reaching a certain `target' region as we considered in Sec.~\ref{sec:ness}.
The full set of rules governing a set of trajectories defines the ensemble by determining the weights of each trajectory as in \eqref{trajdistvec}, and we are often interested in ensemble or average behavior because this is what is usually observed experimentally, although single-molecule studies by now a well-established and important field of study \cite{deniz2008single,miller2017single,astumian2006unreasonable}.

It is critical to appreciate that an individual trajectory generally cannot be considered to be of equilibrium or non-equilibrium character in an intrinsic sense.
(A possible exception is an extremely long trajectory which itself fully embodies all defining criteria of the ensemble. \cite{zuckerman2010statistical})
Generally, it is the \emph{distribution} of trajectories that determines whether a system is in equilibrium and, if not, what ensemble it represents.
Two finite-length trajectories that have the same weight in the equilibrium ensemble might have different weights in a non-equilibrium ensemble.
The trajectory distribution will be determined by the initial phase-point distribution $p_0$ in conjunction with the imposed boundary and thermodynamic conditions such as temperature.

Let us consider equilibrium in the trajectory ensemble picture.
For simplicity, we will assume that our initial phase point distribution is already Boltzmann-distributed: $p_0(\xvec_0) \propto \exp(- H(\xvec_0) / k_B T)$.
As trajectories evolve in time from their initial points, the system will remain in equilibrium if the thermodynamic and boundary conditions remain the same.
Thus, dynamics underlie equilibrium.
We can say dynamics \emph{define} equilibrium through detailed balance: if we count transitions occurring between small volumes around phase points $\xvec_i$ and $\xvec_j$ over any interval of time, the counts $i \to j$ and $j \to i$ will be identical within noise; the same is true for any size volumes in equilibrium. \cite{onsager1931reciprocal,zuckerman2010statistical}
This detailed balance property not only keeps the distribution stationary in time, but it means there are no net flows anywhere in phase space.
Detailed balance further implies there is no net flow along any trajectory-like path -- i.e., the forward and exactly time-reversed trajectories will occur an equal number of times. \cite{crooks1999entropy,astumian2020nonequilibrium}

Note that our discussion here applies to thermal (constant-temperature) equilibrium for systems whose full configurations or phase points may include real-space coordinates and/or chemical degrees of freedom.
That is, the trajectory picture of equilibrium applies: for conformational processes in molecules, such as isomerization or folding; for simple diffusion or diffusion with possibly space-varying `drift' forces; for molecular binding, which may include both translational and conformational processes; for chemical processes involving electronic degrees of freedom such as bond formation and breakage; and for any combination of these whether modeled in full detail or approximately -- so long as there is no implicit addition or removal of energy or particles.
The trajectory picture does not apply for mechanical equilibrium, the balance of forces.

\section{Powerful simulation methodology based on trajectory ensembles}
\label{sec:we}

\subsection{Goals and challenges of computation}
To consider computational strategies, we should first understand the goals of computation.
As we do so, keep in mind a concrete process like protein folding or another spontaneous transition from a metastable state to a more stable one, such as a conformational change in a protein, a change in crystal lattice form, or a re-arrangement of a molecular cluster.
For any of these transitions we might be interested in the following:
\begin{enumerate}[label=(\roman*)]
    \item the ``kinetics'' -- the MFPT or some other measure of rate for the transition
    \item the ``mechanism'' or pathways of the process -- the sequence(s) of states exhibited during the transition
    \item the ``relaxation'' process -- the timescales and mechanisms of describing the transient way the system `settles in' to a steady state
\end{enumerate}

\begin{figure}[h]
    \centering
    \includegraphics[width=12cm]{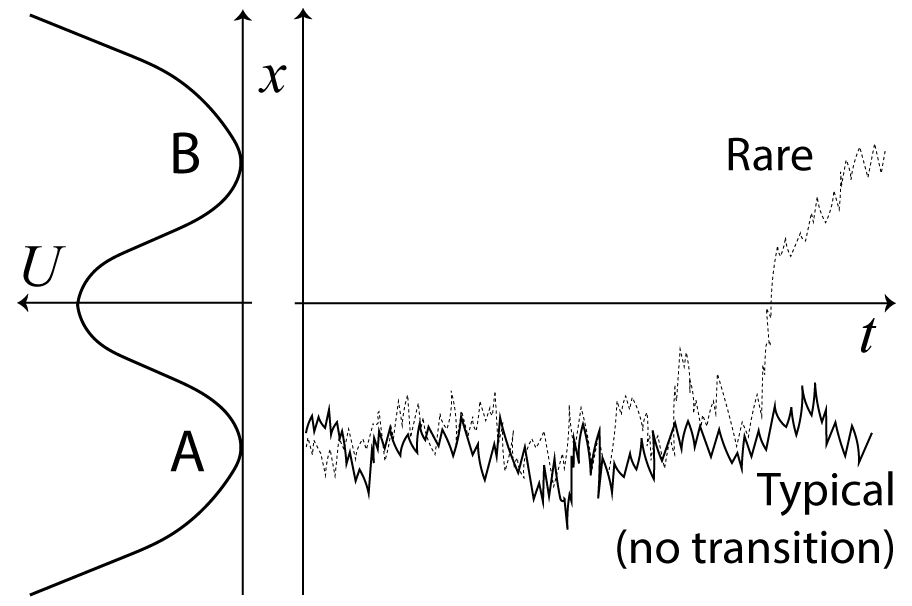}
    \caption{The challenge of rare-event sampling in computation.  Trajectories are initiated in state A, but in challenging systems most will remain in state A (solid trajectory).  Transitions (dotted line) may be extremely unlikely or effectively unobservable in realistic, high-complexity systems such as protein conformational changes.  Hence, typical ``brute force'' simulations can be both wasteful and expensive.}
    \label{fig:challenges}
\end{figure}

We'll first consider a simple, though typically impractical, way to calculate any or all of the above.
As sketched in Fig.\ \ref{fig:challenges}, the naive ``brute force'' implementation would simply be to initiate a large number of trajectories using an initial distribution of interest $p_0$ and wait until all trajectories have made the transition of interest.
From this set of trajectories, we could (i) average their durations to obtain the MFPT or (ii) analyze the states occurring during transitions to quantify the mechanism.\cite{suarez2018pathway}
For (iii) relaxation, we could wait still longer until the spatial/configurational distribution becomes stationary (using `recycling' if studying a constant-temperature NESS) and quantify the relaxation time as well as the mechanism, perhaps via probability shifts that occur.
However, the strategy of waiting for multiple spontaneous transitions will only work for the simplest systems, such as low-dimensional toy models -- see Exercises.

In general, the brute force approach will \emph{not} be practical for complex systems.
And if a system is complicated and directly pertinent to real-world problems, it's likely to be too expensive to permit thorough brute-force simulation.
We can quantify the challenges with a back-of-the-envelope calculation.
For the system of interest, say you can afford a total of $M$ simulations of duration $\tmax$.
This means, roughly, that you can determine the distribution of phase points at any time $t<\tmax$, denoted $p(\xvec,t)$, to a precision of $1/M$ ... and typically you won't have knowledge of behavior beyond $\tmax$.
As a point of reference in biomolecules, current hardware limits $\tmax$ to 1-10 $\mu$s in most systems (and to ms for small systems with extraordinary resources \cite{lindorff2011fast}), whereas most biological phenomena occur on a timescale of at least 100 $\mu$s and more typically on ms - s scales.

\subsection{Efficient simulation via the weighted ensemble approach}

Fortunately, there are now methods \cite{allen2005sampling,faradjian2004computing,van2003novel,warmflash2007umbrella,chong2017path} that can sidestep the $1/M$ limitation just described, and we'll focus on the most straightforward of these, known as the \emph{weighted ensemble} (WE) strategy. \cite{huber1996weighted,zhang2010weighted,zuckerman2017weighted}
WE is a multi-trajectory ``splitting method'' based on a proposal credited to von Neumann \cite{kahn1951estimation} that can provide information on relaxation and steady-state behavior.
WE can provide this information using less \emph{overall} computing than naive simulation, i.e., the product $M \tmax$ is smaller.
It achieves this by re-allocating computing effort (trajectories) away from easy-to-sample regions of phase space toward rarer regions.
WE is also an unbiased method: on average, it exactly recapitulates trajectory ensemble behavior and hence the time-evolution of the spatial distribution $p(\xvec,t)$; \cite{zhang2010weighted} the latter property reflects consistency with the Fokker-Planck equation, \cite{Risken1996,gardiner2009stochastic} which is briefly described in Appendix \ref{app:smol}.

\begin{figure}[h]
    \centering
    \includegraphics[width=15cm]{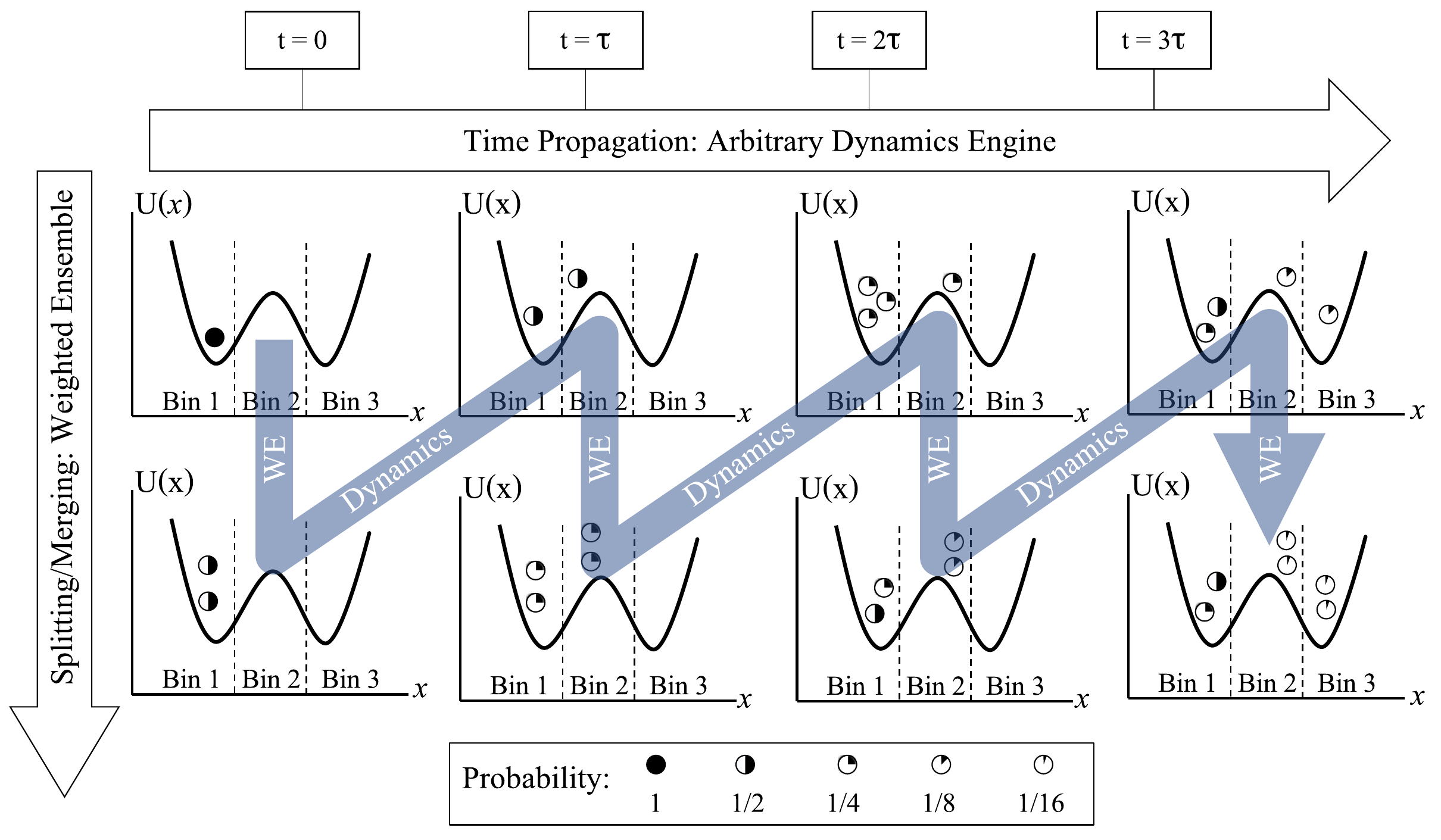}
    \caption{Efficient simulation via the weighted ensemble (WE) method. \cite{donovan2016unbiased}  Phase space is divided into bins, and trajectories are started according to an initial distribution of interest (far left).  Dynamics are run briefly, allowing trajectories to visit other bins, after which the WE steps of ``splitting'' (replication) and ``merging'' (pruning) are performed.  Weights of parent trajectories are shared among children from splitting events, permitting the estimation of very low-probability events.  In this example, a target of two trajectories per bin has been set.
    Donovan et. al, PLOS Computational Biology, 12, 2, 2016; licensed under a Creative Commons Attribution (CC BY) license.
    }
    \label{fig:we}
\end{figure}

WE simulation follows a fairly simple procedure, schematized in Fig.\ \ref{fig:we}, which promotes the presence of trajectories in relatively rare regions of an energy landscape.
In a basic implementation, \cite{huber1996weighted} phase space is divided into non-overlapping bins of the user's construction, and a target number of trajectories per bin is set -- say, 2, for concreteness.
The bins should finely subdivide difficult-to-sample regions such as energy barriers to enable ``statistical ratcheting'' up hills if trajectories are examined frequently enough.
That  is, because short trajectories always have some probability to move uphill in energy, brief unbiased fluctuations can be `captured' for ratcheting and effectively concatenated to study otherwise rare events, sidestepping the $1/M$ limitation.
Trajectories are started at the user's discretion, but let's assume two trajectories are started in a bin of state A, with the goal of sampling transitions to B.

Trajectories in WE are run in parallel for brief intervals of time $\tau$ (with $\mfpt \gg \tau \gg \dt$, where $\dt$ is the simulation time step), then stopped and restarted according to simple probabilistic rules.
In our example, each of the two trajectories is initially given a weight $1/2$ at $t=0$ and the essential idea is to ensure probability moves in an unbiased way, thus preserving the trajectory ensemble behavior and $p(\xvec,t)$.
If a trajectory is found to occupy an otherwise empty bin after one of the $\tau$ intervals, \emph{two} ``child'' trajectories are initiated from the final phase point of the ``parent'' trajectory, and the children each inherit $1/2$ of the parent's weight --- a process called \emph{splitting}.
The two child trajectories in the previously unvisited bin create the ratcheting effect: there is twice the likelihood to explore that region, and to continue to still rarer regions, than if we did not replicate trajectories.
Stochastic dynamics must be used, otherwise child trajectories will evolve identically.

If more than two WE trajectories are found in a bin, pruning (or \emph{merging}) is performed in a pairwise fashion: a random number is generated to select one of an arbitrary pair for continuation with probabilities proportional to their weights, and the selected trajectory absorbs the weight of the other trajectory, which is discontinued.
In this fashion, energy minima do not collect large numbers of trajectories which would add cost to the simulation but provide minimal statistical value.
The processes described, in fact, constitute unbiased statistical \emph{resampling} \cite{zhang2010weighted} -- see Exercises.
In WE, the total trajectory cost is limited to the number of bins multiplied by the number of trajectories per bin and the trajectory length.
This amounts to $M \, \tmax$ in our case, given $M/2$ bins.

\begin{figure}
    \centering
    \includegraphics{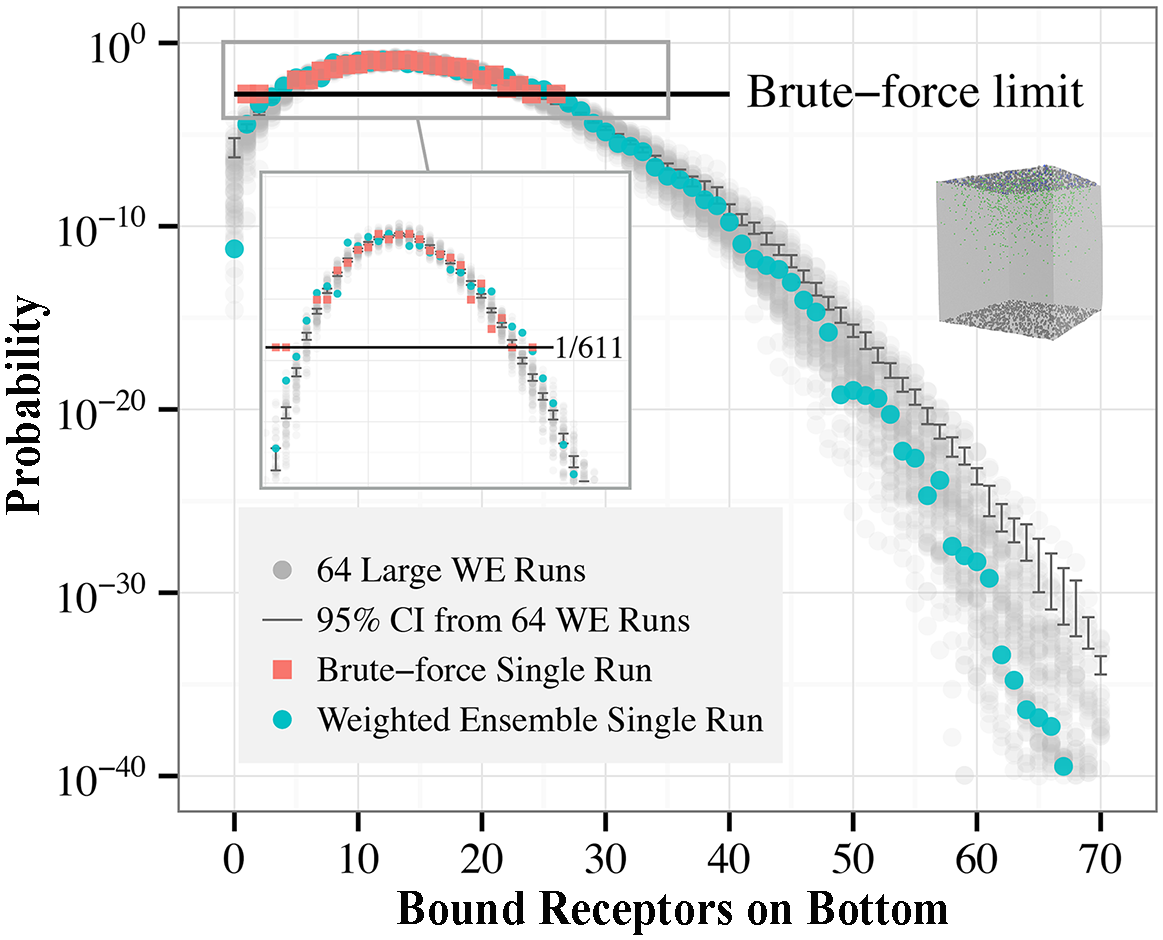}
    \caption{Weighted ensemble simulation of extremely rare diffusion and binding events. \cite{donovan2016unbiased}  Particles are initiated at the top of a three-dimensional box (upper right inset) and allowed to diffuse without bias.  Any particles that reach the bottom surface of the box can bind to receptors located there.  The graph shows the probability distribution of bound receptors after a short time interval -- i.e., the likelihood of different outcomes that would result from a single brute-force diffusion simulation.  WE enables sampling deep into the tails of the distribution because more trajectories are allotted to rarer outcomes, whereas an equivalent amount of ``brute force'' sampling cannot detect events rarer than the reciprocal of the number of trajectories, as shown by solid horizontal lines.  WE simulations used simulation time equivalent to 611 brute force trajectories, as indicated in the left inset.  The grey dots represent independent WE runs (of which green is a representative) and solid vertical bars give the confidence interval based on the grey data -- which appears to be skewed upward because of the logarithmic scale.
    Donovan et. al, PLOS Computational Biology, 12, 2, 2016; licensed under a Creative Commons Attribution (CC BY) license.}
    \label{fig:wetoy}
\end{figure}

Although the total simulation cost is bounded by $M \, \tmax$ (plus overhead for splitting/merging), events \emph{much} rarer than $1/M$ can be seen because of the splitting procedure.
Indeed, exponentially rare processes are elicited as WE produces an unbiased estimate of the trajectory ensemble and $p(\xvec,t)$.
A dramatic example is shown in Fig.\ \ref{fig:wetoy} for diffusion and binding in a 3D box, where the distribution of possible binding outcomes extends \emph{tens of orders of magnitude below what standard simulation provides.}
For monitoring the transient time evolution of a system, WE is almost like a ``magic bullet.''

Obtaining the MFPT from WE simulation is more challenging than characterizing $p(\xvec,t)$ in many cases.
To use the Hill relation \eqref{hill}, the system must relax to steady state and this relaxation is \emph{not} accelerated by WE for the very reason it is so successful in characterizing $p(\xvec, t)$, i.e., because it is unbiased.
To see this more concretely, let $\tss$ be the average time required for a given system to relax to steady state.
Then, because WE runs $M$ copies of the system, the total cost for observing a WE simulation relax to steady state is $\sim M \, \tss$, which will be prohibitive in some though not all systems. \cite{adhikari2018computational}
Even when $M \, \tss$ is a prohibitive cost, the MFPT can be obtained from transient data ($t < \tss$) available in WE simulation: although the details are beyond the scope of this discussion, the idea is to use much finer-grained and faster-relaxing bins (than were used to run the WE simulation) in a quasi-Markov approximation. \cite{copperman2020accelerated}
Below, we apply WE directly for MFPT calculation in a simple system.

Like any advanced computational method, WE has its subtleties and limitations.
Most important are correlations.
Although WE trajectories are independent (non-interacting), exactly as assumed in the trajectory-ensemble definitions, correlations arise in the overall WE protocol due to the splitting and merging steps.
After all, when a trajectory is ``split,'' by construction the child trajectories are identical until the split point.
Therefore assessing statistical uncertainty in WE estimates requires great care, even though the method is unbiased. \cite{mostofian2019statistical}

\subsection{Applying the weighted ensemble to a simple model}
\label{sec:wedbl}

To illustrate the power and validity of the weighted ensemble method, we employ it to estimate the transition rate over a high energy barrier in a simple system.
We use the WESTPA implementation\cite{zwier2015westpa} of WE and apply it to a simple 1D double-well potential under overdamped Langevin dynamics \eqref{overdamped} with parameters chosen to approximate the behavior of a small molecule in water. 
We assume a mass of 100 u, temperature $T = 300$K, a barrier height of $10 k_B T$, and a friction coefficient $\gamma = 24.94 \mathrm{ps}^{-1}$ which is reasonable for water and corresponds to a diffusion constant of $10^{-6} \mbox{cm}^2/\mbox{sec}$. 
The simulation is run with a timestep of $3$ ps, and all simulation code is available on Github \cite{russo2020github}.

\begin{figure}
    \centering
    \includegraphics[width=0.45\linewidth]{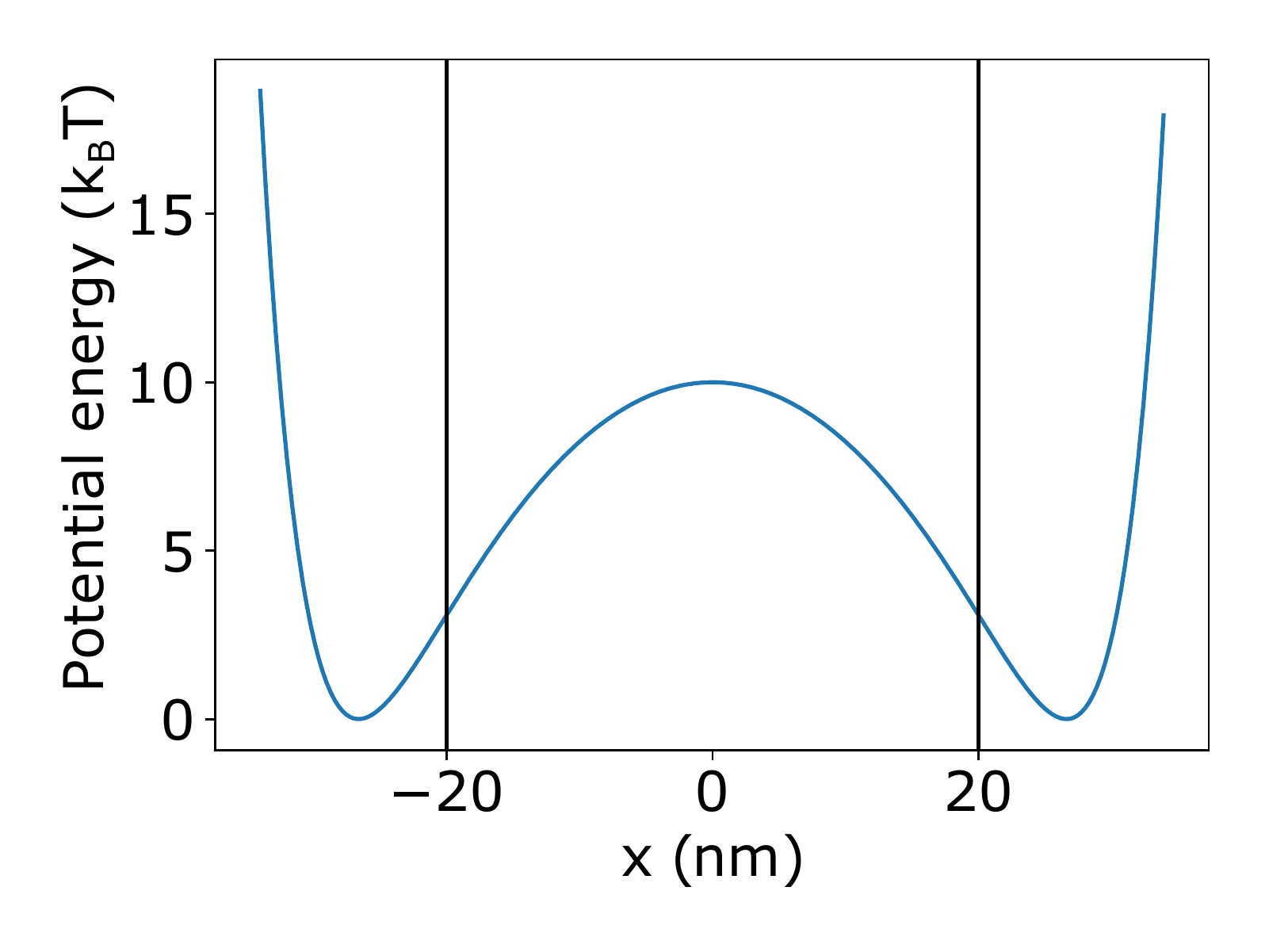}
    \includegraphics[width=0.52\linewidth]{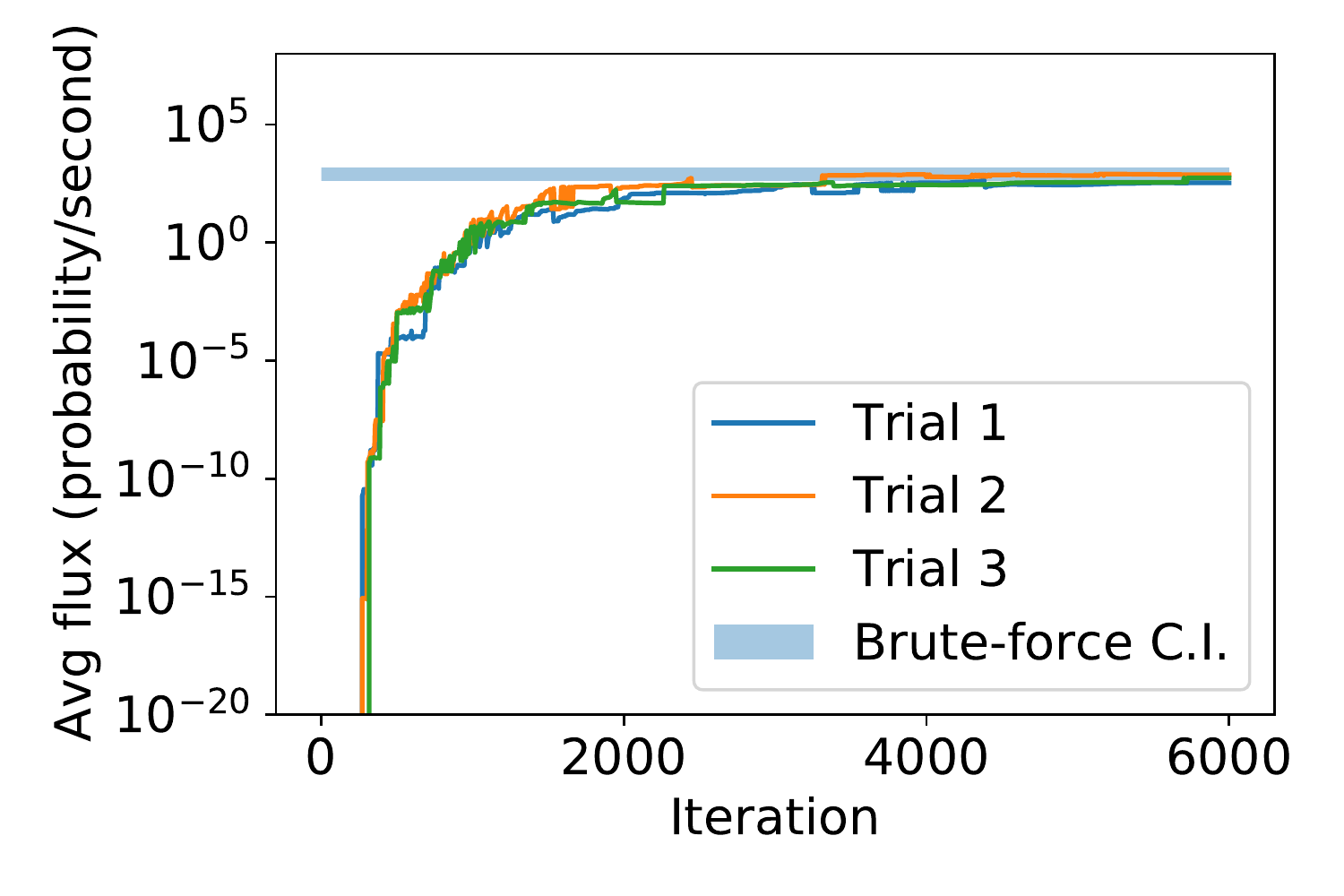}
    \caption{Weighted ensemble estimation of the rate of a rare event: high-barrier crossing.  (a): Potential energy function used for double-well simulation with 10 $k_B T$ barrier and state boundaries indicated by the vertical black lines.
    (b): Average flux into the left basin state for simulations started from the right basin, as computed from three independent weighted ensemble simulations (colored lines).  The average flux estimates the inverse MFPT by Eq.\ \eqref{hill}, yielding $\sim1$ ms.  For reference, an independent estimate of the flux is computed using a very long ``brute force'' simulation (horizontal line). 
    The brute force confidence interval (C.I.) is shown as a blue shaded region, which is $\pm$ twice the standard error of the mean based on 11 transitions. 
    }
    \label{fig:we_dbl_well}
\end{figure}

The WE simulation is set up with walkers beginning in the rightmost basin, 
and with the two basin macrostates defined as $x > 20$ nm and $x<-20$ nm, 
as shown in Fig.~\ref{fig:we_dbl_well}.  
Twenty uniform bins of width 2 nm uniformly span 
from $x=$-20.0 to 20.0 nm, 
with two additional bins on either end reaching to $\pm \infty$. 
The WE simulation is run with a resampling time of $\tau = $ 60 ps 
and a target count of 10 trajectories per bin so roughly 200 trajectories will run during each $\tau$.
Walkers that reach the left basin are ``recycled'' and restarted from $x=20$ nm to generate a non-equilibrium steady state and exploit the Hill relation \eqref{hill}.

To quantify the effectiveness of WE simulation for this case, we can compare the cost for computing the rate constant (i.e., flux or reciprocal MFPT) from WE simulation to brute-force simulation of overdamped Langevin dynamics.
Note from Fig.\ \ref{fig:we_dbl_well} that WE simulations reach steady values after $\sim$3,000 iterations, which corresponds to $\sim12$ million steps of total simulation (for a single WE run, accounting for all $\sim$ 200 trajectories) or a total of 36 $\mu$s of simulated time. 
Also from Fig.\ \ref{fig:we_dbl_well} and the Hill relation \eqref{hill}, the MFPT is $\sim 1000 \mu$s. 
Thus, we see that WE simulation has generated the \emph{average} first-passage time using an overall amount of computing that is only a small fraction ($\sim0.04$) of the time needed to yield a \emph{single} transition event via direct simulation, let alone to generate a reliable MFPT estimate from multiple events.

\section{Concluding Discussion}

The trajectory arguably is the most fundamental object in classical statistical mechanics, particularly for non-equilibrium phenomena, and this article has attempted to connect trajectory physics with more familiar topics in the traditional physics curriculum.
By focusing in depth on the simplest possible example -- diffusion -- we have been able to formalize and visualize the probabilistic/ensemble picture and connect it with simpler spatial distributions.
We have further been able to connect these ensembles with observable populations, kinetics, and thermodynamic states, as well as understand a modern, practical path-sampling approach.

\begin{figure}[h]
    \centering
    \includegraphics[width=15cm]{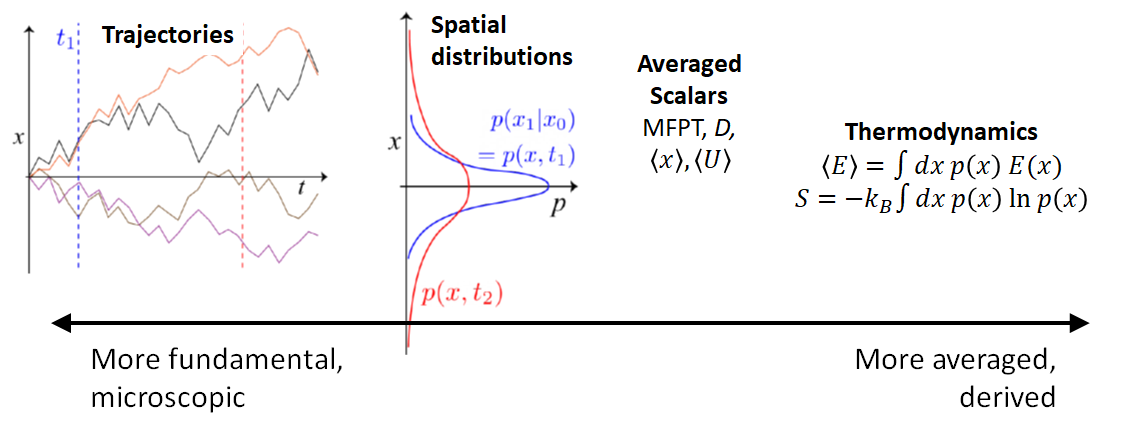}
    \caption{From the fundamental ensemble of trajectories to more averaged observables.  Because trajectories embody the dynamics that fully specifies a system, they are the most fundamental.  Averaging or analysis can be performed at fixed time points, including the the $t \to \infty$ stationary point.
    Quantities that can be calculated include the phase-space distribution $p(x,t)$, the mean first-passage time (MFPT), diffusion constant ($D$), average coordinates or properties (e.g., $\langle x \rangle, \langle U \rangle$) or system-wide thermodynamic properties, in or out of equilibrium.  Although simple diffusive trajectories are pictured, the same principles apply in the case of non-zero forces.}
    \label{fig:averaging}
\end{figure}

A key lesson is that theoretical physics can view a given process at different levels of ``magnification,'' from most microscopic to most averaged (Fig.\ \ref{fig:averaging}).
Trajectories are the most detailed and encompass all system coordinates at all times -- which is usually too much to grasp.
Trajectories can be averaged spatially at fixed times to yield more familiar probability distributions.
Trajectory flows across surfaces of interest can also be averaged to yield probability fluxes: in equilibrium, all such fluxes are zero, whereas in transient regimes or non-equilibrium steady states (NESS's), such flows provide key information.
Notably the Hill relation \eqref{hill} yields the mean first-passage time (MFPT) from the flux in an appropriate NESS, and furthermore, conditions on reversibility can be derived from flux arguments (Appendix \ref{app:revers}).
Finally, averaging -- i.e., integrating -- over spatial distributions can yield observable thermodynamic information on state populations \cite{reif2009fundamentals,zuckerman2017weighted}; see also entropy production and fluctuation relations. \cite{bochkov1981nonlinear,seifert2012stochastic,jarzynski1997nonequilibrium,crooks1999entropy}

This report has only given a taste of the value of the trajectory picture, which goes much further.
Trajectory ideas, for example, are used to develop the Jarzynski relation. \cite{jarzynski1997nonequilibrium,jarzynski1998equilibrium,crooks1998nonequilibrium}
They provide a direct connection with the path-integral formulation of quantum mechanics. \cite{Sakurai1985modern}
Trajectories offer a unique window into the often mis-understood issue of ``reversibility'' \cite{bhatt2011revers} -- see Appendix \ref{app:revers}.
Not surprisingly, trajectories and their applications are still an area of active research. \cite{seifert2012stochastic,thiede2019galerkin,aristoff2020optimizing,rotskoff2019dynamical,russo2020iterative,adhikari2018computational}

\section{Exercises}
\label{sec:exercises}
\begin{enumerate}
    \item Confirm by differentiation that \eqref{pnposn} is the exact solution to the diffusion equation \eqref{diffpde}, after setting $x_n = x$ and $n = t/\dt$.  Note that $t$ occurs both in the prefactor \emph{and} the exponent, so differentiation requires the product rule.
    
    \item Time-discretization generally introduces an error into dynamics computed via \eqref{dxoverd} and \eqref{dxdet}.  Explain why there is an error and how it might be mitigated in computer simulation.  For what special case is there no error even if $f\neq 0$?
    
    \item \label{ex:dbl} Implement overdamped dynamics simulation \eqref{dxoverd} of the double-well system specified in Sec.\ \ref{sec:wedbl}.  Calculate the MFPT of the system for a range of barrier heights, starting with a low barrier, by simple averaging of $\sim$10 observed first-passage times.  Compare these values to the expected Arrhenius behavior. \cite{zuckerman2010statistical}
    
    \item Using the ODLD module of the WESTPA implementation of weighted ensemble, implement a \emph{triple} well system.  Consider the left-most basin to be the initial state (A) and the right-most basin the target (B).  Examine the relaxation of the probability into the target state as a function of time.  For cleanest data, average over multiple WE runs.  Vary the depth of the middle well and explain the observed behavior.
    
    \item Write down the trajectory probability, the analog of \eqref{trajdist}, for a system with constant force, sometimes called simple drift.  Explain in words the meaning of the distribution.  If you can, integrate out intermediate time points to show the behavior remains Gaussian with constant drift.
    
    \item For a simple diffusive system described by \eqref{trajdist}, obtain the distribution for $x_3$ by a suitable integration of \eqref{ptwoposn}.
    
    \item Write down the equations that define (i) a steady state and (ii) equilibrium for a discrete-state system in terms of steady probabilities $p_i$ and state-to-state transition probabilities $\tij$ for some fixed time interval.  Note that equilibrium is defined by detailed balance.  Show that detailed balance implies steady state but not the reverse.  A counter-example suffices to disprove a hypothesis.
    
    
    \item By studying the theory underlying weighted ensemble, \cite{zhang2010weighted} explain in statistical terms why the ``resampling'' procedure for ``merging'' trajectories does not bias the time-evolving probability distribution $p(\xvec,t)$.
    
    \item Write pseudocode for a weighted ensemble simulation of an arbitrary system with pre-defined bins.  If you are ambitious, implement your pseudocode for 1D overdamped dynamics in the double-well system in Exercise \ref{ex:dbl}.
    
    \item Understand the continuity equation \eqref{cont} by integrating it over an interval in $x$ from $a$ to $b$.
    Integrating the probability density over this region gives the total probability in it.  How does this probability change in time based on the current and why does the result make sense?  Remember the one-dimensional current is defined to be positive in the right-ward direction.
    \label{ex:cont}
    
    \item Show that stationary distribution of the Smoluchowski equation \eqref{smol}, i.e., when $\partial p / \partial t = 0$, is the expected equilibrium distribution based on the Boltzmann factor.
\end{enumerate}

\begin{acknowledgments}
The authors are grateful for support from the National Science Foundation under grant MCB 1715823 and from the National Institutes of Health under grant GM115805.
We very much appreciate helpful discussions with Jeremy Copperman and Ernesto Suarez.

\end{acknowledgments}


\begin{thebibliography}{99}

    \bibitem{onsager1931reciprocal}
    Lars Onsager.
    \newblock Reciprocal relations in irreversible processes. {I.}
    \newblock {\em Physical review}, 37(4):1505--1512, 1931.
    
    \bibitem{vanden2010transition}
    Eric Vanden-Eijnden et~al.
    \newblock Transition-path theory and path-finding algorithms for the study of
      rare events.
    \newblock {\em Annual review of physical chemistry}, 61:391--420, 2010.
    
    \bibitem{elber2020molecular}
    Ron Elber, Dmitrii~E Makarov, and Henri Orland.
    \newblock {\em Molecular kinetics in condensed phases: Theory, simulation, and
      analysis}.
    \newblock John Wiley \& Sons, 2020.
    
    \bibitem{zuckerman2010statistical}
    Daniel~M Zuckerman.
    \newblock {\em Statistical physics of biomolecules: {A}n introduction}.
    \newblock CRC Press, Boca Raton, FL, 2010.
    
    \bibitem{giancoli2008physics}
    Douglas~C Giancoli.
    \newblock {\em Physics for scientists and engineers with modern physics}.
    \newblock Pearson Education, 2008.
    
    \bibitem{reif2009fundamentals}
    Frederick Reif.
    \newblock {\em Fundamentals of statistical and thermal physics}.
    \newblock McGraw-Hill, New York, 1965.
    
    \bibitem{seifert2012stochastic}
    Udo Seifert.
    \newblock Stochastic thermodynamics, fluctuation theorems and molecular
      machines.
    \newblock {\em Reports on Progress in Physics}, 75(12):126001, 2012.
    
    \bibitem{Zuckerman2020keybio}
    Daniel~M Zuckerman.
    \newblock Key biology you should have learned in physics class: Using ideal-gas
      mixtures to understand biomolecular machines.
    \newblock {\em Am. J. Phys.}, 88(3):182--193, March 2020.
    
    \bibitem{lindorff2011fast}
    Kresten Lindorff-Larsen, Stefano Piana, Ron~O Dror, and David~E Shaw.
    \newblock How fast-folding proteins fold.
    \newblock {\em Science}, 334(6055):517--520, 2011.
    
    \bibitem{fersht1999structure}
    Alan Fersht et~al.
    \newblock {\em Structure and mechanism in protein science: a guide to enzyme
      catalysis and protein folding}.
    \newblock Macmillan, 1999.
    
    \bibitem{juraszek2006sampling}
    J~Juraszek and PG~Bolhuis.
    \newblock Sampling the multiple folding mechanisms of trp-cage in explicit
      solvent.
    \newblock {\em Proceedings of the National Academy of Sciences},
      103(43):15859--15864, 2006.
    
    \bibitem{mobley2017predicting}
    David~L Mobley and Michael~K Gilson.
    \newblock Predicting binding free energies: frontiers and benchmarks.
    \newblock {\em Annual review of biophysics}, 46:531--558, 2017.
    
    \bibitem{gunasekaran2004allostery}
    K~Gunasekaran, Buyong Ma, and Ruth Nussinov.
    \newblock Is allostery an intrinsic property of all dynamic proteins?
    \newblock {\em Proteins: Structure, Function, and Bioinformatics},
      57(3):433--443, 2004.
    
    \bibitem{zwier2016efficient}
    Matthew~C Zwier, Adam~J Pratt, Joshua~L Adelman, Joseph~W Kaus, Daniel~M
      Zuckerman, and Lillian~T Chong.
    \newblock Efficient atomistic simulation of pathways and calculation of rate
      constants for a protein--peptide binding process: application to the mdm2
      protein and an intrinsically disordered p53 peptide.
    \newblock {\em The journal of physical chemistry letters}, 7(17):3440--3445,
      2016.
    
    \bibitem{copeland2006drug}
    Robert~A Copeland, David~L Pompliano, and Thomas~D Meek.
    \newblock Drug--target residence time and its implications for lead
      optimization.
    \newblock {\em Nature reviews Drug discovery}, 5(9):730--739, 2006.
    
    \bibitem{zuckerman2017weighted}
    Daniel~M Zuckerman and Lillian~T Chong.
    \newblock Weighted ensemble simulation: review of methodology, applications,
      and software.
    \newblock {\em Annual review of biophysics}, 46:43--57, 2017.
    
    \bibitem{pratt1986statistical}
    Lawrence~R Pratt.
    \newblock A statistical method for identifying transition states in high
      dimensional problems.
    \newblock {\em The Journal of chemical physics}, 85(9):5045--5048, 1986.
    
    \bibitem{dellago1998transition}
    Christoph Dellago, Peter~G Bolhuis, F{\'e}lix~S Csajka, and David Chandler.
    \newblock Transition path sampling and the calculation of rate constants.
    \newblock {\em The Journal of chemical physics}, 108(5):1964--1977, 1998.
    
    \bibitem{zuckerman1999dynamic}
    Daniel~M Zuckerman and Thomas~B Woolf.
    \newblock Dynamic reaction paths and rates through importance-sampled
      stochastic dynamics.
    \newblock {\em The Journal of chemical physics}, 111(21):9475--9484, 1999.
    
    \bibitem{bier1999intrawell}
    Martin Bier, Imre Der{\'e}nyi, Marcin Kostur, and R~Dean Astumian.
    \newblock Intrawell relaxation of overdamped brownian particles.
    \newblock {\em Physical Review E}, 59(6):6422-6432, 1999.
    
    \bibitem{astumian2006unreasonable}
    R~Dean Astumian.
    \newblock The unreasonable effectiveness of equilibrium theory for interpreting
      nonequilibrium experiments.
    \newblock {\em American journal of physics}, 74(8):683--688, 2006.
    
    \bibitem{astumian2010thermodynamics}
    R~Dean Astumian.
    \newblock Thermodynamics and kinetics of molecular motors.
    \newblock {\em Biophysical journal}, 98(11):2401--2409, 2010.
    
    \bibitem{astumian2020nonequilibrium}
    R~Dean Astumian, Cristian Pezzato, Yuanning Feng, Yunyan Qiu, Paul~R McGonigal,
      Chuyang Cheng, and J~Fraser Stoddart.
    \newblock Non-equilibrium kinetics and trajectory thermodynamics of synthetic
      molecular pumps.
    \newblock {\em Materials Chemistry Frontiers}, 4(5):1304--1314, 2020.
    
    \bibitem{hill1982linear}
    Terrell~L Hill.
    \newblock The linear {Onsager} coefficients for biochemical kinetic diagrams as
      equilibrium one-way cycle fluxes.
    \newblock {\em Nature}, 299(5878):84--86, 1982.
    
    \bibitem{hill2004free}
    Terrell~L Hill.
    \newblock {\em Free energy transduction and biochemical cycle kinetics}.
    \newblock Dover, Mineola, NY, 2004.
    
    \bibitem{ghosh2006teaching}
    Kingshuk Ghosh, Ken~A Dill, Mandar~M Inamdar, Effrosyni Seitaridou, and Rob
      Phillips.
    \newblock Teaching the principles of statistical dynamics.
    \newblock {\em American journal of physics}, 74(2):123--133, 2006.
    
    \bibitem{presse2013principles}
    Steve Press{\'e}, Kingshuk Ghosh, Julian Lee, and Ken~A Dill.
    \newblock Principles of maximum entropy and maximum caliber in statistical
      physics.
    \newblock {\em Reviews of Modern Physics}, 85(3):1115-1141, 2013.
    
    \bibitem{ghosh2020maximum}
    Kingshuk Ghosh, Purushottam~D Dixit, Luca Agozzino, and Ken~A Dill.
    \newblock The maximum caliber variational principle for nonequilibria.
    \newblock {\em Annual review of physical chemistry}, 71:213--238, 2020.
    
    \bibitem{swendsen2008explaining}
    Robert~H Swendsen.
    \newblock Explaining irreversibility.
    \newblock {\em American Journal of Physics}, 76(7):643--648, 2008.
    
    \bibitem{chandrasekhar1943stochastic}
    Subrahmanyan Chandrasekhar.
    \newblock Stochastic problems in physics and astronomy.
    \newblock {\em Reviews of modern physics}, 15(1):1-89, 1943.
    
    \bibitem{chong2017path}
    Lillian~T Chong, Ali~S Saglam, and Daniel~M Zuckerman.
    \newblock Path-sampling strategies for simulating rare events in biomolecular
      systems.
    \newblock {\em Current opinion in structural biology}, 43:88--94, 2017.
    
    \bibitem{vanKampen1992stochastic}
    Nicolaas~Godfried Van~Kampen.
    \newblock {\em Stochastic processes in physics and chemistry}, volume~1.
    \newblock Elsevier, 1992.
    
    \bibitem{onsager1953fluctuations}
    Lars Onsager and Stefan Machlup.
    \newblock Fluctuations and irreversible processes.
    \newblock {\em Physical Review}, 91(6):1505-1512, 1953.
    
    \bibitem{Sakurai1985modern}
    J~J Sakurai.
    \newblock {\em Modern Quantum Mechanics}.
    \newblock Addison Wesley, 1985.
    
    \bibitem{gardiner2009stochastic}
    Crispin Gardiner.
    \newblock {\em Stochastic methods}, volume~4.
    \newblock Springer Berlin, 2009.
    
    \bibitem{Risken1996}
    Hannes Risken and Till Frank.
    \newblock {\em The Fokker-Planck Equation: Methods of Solution and
      Applications}, volume~18.
    \newblock Springer Science \& Business Media, 1996.
    
    \bibitem{redner2001guide}
    Sidney Redner.
    \newblock {\em A guide to first-passage processes}.
    \newblock Cambridge University Press, 2001.
    
    \bibitem{chandler1987introduction}
    David Chandler.
    \newblock {\em Introduction to modern statistical mechanics}.
    \newblock Oxford University Press, Oxford, UK, 1987.
    
    \bibitem{chodera2011robust}
    John~D Chodera, Phillip~J Elms, William~C Swope, Jan-Hendrik Prinz, Susan
      Marqusee, Carlos Bustamante, Frank No{\'e}, and Vijay~S Pande.
    \newblock A robust approach to estimating rates from time-correlation
      functions.
    \newblock {\em arXiv preprint arXiv:1108.2304}, 2011.
    
    \bibitem{deniz2008single}
    Ashok~A Deniz, Samrat Mukhopadhyay, and Edward~A Lemke.
    \newblock Single-molecule biophysics: at the interface of biology, physics and
      chemistry.
    \newblock {\em Journal of the Royal Society Interface}, 5(18):15--45, 2008.
    
    \bibitem{miller2017single}
    Helen Miller, Zhaokun Zhou, Jack Shepherd, Adam~JM Wollman, and Mark~C Leake.
    \newblock Single-molecule techniques in biophysics: a review of the progress in
      methods and applications.
    \newblock {\em Reports on Progress in Physics}, 81(2):024601, 2017.
    
    \bibitem{crooks1999entropy}
    Gavin~E Crooks.
    \newblock Entropy production fluctuation theorem and the nonequilibrium work
      relation for free energy differences.
    \newblock {\em Physical Review E}, 60(3):2721-2726, 1999.
    
    \bibitem{suarez2018pathway}
    Ernesto Su{\'a}rez and Daniel~M Zuckerman.
    \newblock Pathway histogram analysis of trajectories: A general strategy for
      quantification of molecular mechanisms.
    \newblock {\em arXiv preprint arXiv:1810.10514}, 2018.
    
    \bibitem{allen2005sampling}
    Rosalind~J Allen, Patrick~B Warren, and Pieter~Rein Ten~Wolde.
    \newblock Sampling rare switching events in biochemical networks.
    \newblock {\em Physical review letters}, 94(1):018104, 2005.
    
    \bibitem{faradjian2004computing}
    Anton~K Faradjian and Ron Elber.
    \newblock Computing time scales from reaction coordinates by milestoning.
    \newblock {\em The Journal of chemical physics}, 120(23):10880--10889, 2004.
    
    \bibitem{van2003novel}
    Titus~S van Erp, Daniele Moroni, and Peter~G Bolhuis.
    \newblock A novel path sampling method for the calculation of rate constants.
    \newblock {\em The Journal of chemical physics}, 118(17):7762--7774, 2003.
    
    \bibitem{warmflash2007umbrella}
    Aryeh Warmflash, Prabhakar Bhimalapuram, and Aaron~R Dinner.
    \newblock Umbrella sampling for nonequilibrium processes.
    \newblock {\em The Journal of chemical physics}, 127(15):114109, 2007.
    
    \bibitem{huber1996weighted}
    Gary~A Huber and Sangtae Kim.
    \newblock Weighted-ensemble brownian dynamics simulations for protein
      association reactions.
    \newblock {\em Biophysical Journal}, 70(1):97, 1996.
    
    \bibitem{zhang2010weighted}
    Bin~W Zhang, David Jasnow, and Daniel~M Zuckerman.
    \newblock The “weighted ensemble” path sampling method is statistically
      exact for a broad class of stochastic processes and binning procedures.
    \newblock {\em The Journal of Chemical Physics}, 132(5):054107, 2010.
    
    \bibitem{kahn1951estimation}
    Herman Kahn and Theodore~E Harris.
    \newblock Estimation of particle transmission by random sampling.
    \newblock {\em National Bureau of Standards applied mathematics series},
      12:27--30, 1951.
    
    \bibitem{donovan2016unbiased}
    Rory~M Donovan, Jose-Juan Tapia, Devin~P Sullivan, James~R Faeder, Robert~F
      Murphy, Markus Dittrich, and Daniel~M Zuckerman.
    \newblock Unbiased rare event sampling in spatial stochastic systems biology
      models using a weighted ensemble of trajectories.
    \newblock {\em PLoS computational biology}, 12(2), 2016.
    
    \bibitem{adhikari2018computational}
    Upendra Adhikari, Barmak Mostofian, Jeremy Copperman, Sundar~Raman Subramanian,
      Andrew~A Petersen, and Daniel~M Zuckerman.
    \newblock Computational estimation of microsecond to second atomistic folding
      times.
    \newblock {\em Journal of the American Chemical Society}, 141(16):6519--6526,
      2019.
    
    \bibitem{copperman2020accelerated}
    Jeremy Copperman and Daniel~M Zuckerman.
    \newblock Accelerated estimation of long-timescale kinetics from weighted
      ensemble simulation via non-markovian “microbin” analysis.
    \newblock {\em Journal of Chemical Theory and Computation}, 16(11):6763--6775,
      2020.
    
    \bibitem{mostofian2019statistical}
    Barmak Mostofian and Daniel~M Zuckerman.
    \newblock Statistical uncertainty analysis for small-sample, high log-variance
      data: cautions for bootstrapping and bayesian bootstrapping.
    \newblock {\em Journal of chemical theory and computation}, 15(6):3499--3509,
      2019.
    
    \bibitem{zwier2015westpa}
    Matthew~C. Zwier, Joshua~L. Adelman, Joseph~W. Kaus, Adam~J. Pratt, Kim~F.
      Wong, Nicholas~B. Rego, Ernesto Su{\'{a}}rez, Steven Lettieri, David~W. Wang,
      Michael Grabe, Daniel~M. Zuckerman, and Lillian~T. Chong.
    \newblock {WESTPA}: An interoperable, highly scalable software package for
      weighted ensemble simulation and analysis.
    \newblock {\em Journal of Chemical Theory and Computation}, 11(2):800--809,
      January 2015.
    
    \bibitem{russo2020github}
    John Russo.
    \newblock jdrusso/doublewell v1.1.
    \newblock {\em Zenodo}, https://doi.org/10.5281/zenodo.4706088, April 2021.
    
    \bibitem{bochkov1981nonlinear}
    GN~Bochkov and Yu~E Kuzovlev.
    \newblock Nonlinear fluctuation-dissipation relations and stochastic models in
      nonequilibrium thermodynamics: I. generalized fluctuation-dissipation
      theorem.
    \newblock {\em Physica A: Statistical Mechanics and its Applications},
      106(3):443--479, 1981.
    
    \bibitem{jarzynski1997nonequilibrium}
    Christopher Jarzynski.
    \newblock Nonequilibrium equality for free energy differences.
    \newblock {\em Physical Review Letters}, 78(14):2690-2693, 1997.
    
    \bibitem{jarzynski1998equilibrium}
    C~Jarzynski.
    \newblock Equilibrium free energies from nonequilibrium processes.
    \newblock {\em Acta Physica Polonica. Series B}, 29(6):1609--1622, 1998.
    
    \bibitem{crooks1998nonequilibrium}
    Gavin~E Crooks.
    \newblock Nonequilibrium measurements of free energy differences for
      microscopically reversible markovian systems.
    \newblock {\em Journal of Statistical Physics}, 90(5-6):1481--1487, 1998.
    
    \bibitem{bhatt2011revers}
    Divesh Bhatt and Daniel~M Zuckerman.
    \newblock Beyond microscopic reversibility: Are observable nonequilibrium
      processes precisely reversible?
    \newblock {\em Journal of chemical theory and computation}, 7(8):2520--2527,
      2011.
    
    \bibitem{thiede2019galerkin}
    Erik~H Thiede, Dimitrios Giannakis, Aaron~R Dinner, and Jonathan Weare.
    \newblock Galerkin approximation of dynamical quantities using trajectory data.
    \newblock {\em The Journal of chemical physics}, 150(24):244111, 2019.
    
    \bibitem{aristoff2020optimizing}
    David Aristoff and Daniel~M Zuckerman.
    \newblock Optimizing weighted ensemble sampling of steady states.
    \newblock {\em Multiscale Modeling \& Simulation}, 18(2):646--673, 2020.
    
    \bibitem{rotskoff2019dynamical}
    Grant~M Rotskoff and Eric Vanden-Eijnden.
    \newblock Dynamical computation of the density of states and bayes factors
      using nonequilibrium importance sampling.
    \newblock {\em Physical review letters}, 122(15):150602, 2019.
    
    \bibitem{russo2020iterative}
    John~D Russo, Jeremy Copperman, and Daniel~M Zuckerman.
    \newblock Iterative trajectory reweighting for estimation of equilibrium and
      non-equilibrium observables.
    \newblock {\em arXiv preprint arXiv:2006.09451}, 2020.
    
\end{thebibliography}

\appendix
\section{The Fokker-Planck picture and Smoluchowski equation in one dimension}
\label{app:smol}

The Fokker-Planck and related equations \cite{Risken1996,gardiner2009stochastic} are essential to understanding non-equilibrium statistical mechanics.
These equations generalize the diffusion equation \eqref{diffpde} but they perform essentially the same role: they quantify the way a spatial and/or configurational distribution changes over time based on a given energy landscape.
The key point is that this is a very general concept that applies not only to center-of-mass diffusive motion but also to configurational motions internal to a molecule or system.
For example, if a protein is started in a certain configuration, where is it likely to be later?
The distribution $p(\xvec,t)$ quantifies the distribution of configurations $\xvec$ at any time $t$.

Here we focus on the Smoluchowski equation, which is the Fokker-Planck equation specific for the overdamped, non-inertial dynamics \eqref{overdamped} studied above.
The Smoluchowski equation is easiest to grasp starting from the continuity equation, given by
\begin{equation}
    \frac{\partial p}{\partial t} = -\frac{\partial J}{\partial x} \;,
    \label{cont}
\end{equation}
in one dimension, where $p = p(x,t)$ is the probability density at time $t$ and $J=J(x,t)$ is the probability current -- i.e., the (average) probability per unit time moving in the $+\!x$ direction.
Note that this is the average over trajectories moving in both directions, so it is the \emph{net} current.
The continuity equation simply ensures that the change of probability in any region is the difference between incoming and outgoing probability.
For students who are new to the continuity equation, Exercise \ref{ex:cont} will clarify its meaning.

To complete the Smoluchowski equation, we need the current corresponding to overdamped dynamics \eqref{overdamped}.
As noted above, overdamped dynamics includes both (simple) diffusion and ``drift'' (motion due to force).
From \eqref{diffpde}, we already can infer that the diffusive current is $-D \, \partial p/\partial x$, which is Fick's law indicating that particles/probability will diffuse down their gradients in a linear fashion on average.
When a force is present, the governing dynamics \eqref{overdamped} indicates that there is also motion linearly proportional to the force, leading to a total current
\begin{equation}
    J(x,t) = -D \frac{\partial p}{\partial x} + \frac{D}{k_BT} f(x) \, p(x,t) \; ,
    \label{joverd}
\end{equation}
where we have assumed $D = k_B T/ m \gamma$ is constant in space.

We obtain the full Smoluchowski equation in one-dimension for fixed $D$ by substituting the current \eqref{joverd} into the continuity equation \eqref{cont}, yielding
\begin{equation}
        \frac{\partial p}{\partial t} = D \frac{\partial^2 p}{\partial x^2} - \frac{D}{k_BT} \frac{\partial }{\partial x} f(x) \, p(x,t) \; .
        \label{smol}
\end{equation}
The diffusion equation has been augmented by a term dependent on the force.
Equation \eqref{smol} can be solved to find the steady-state behavior of $p$ both out of or in equilibrium (see Exercises) or to follow the time-dependent behavior as the distribution $p$ relaxes toward its limiting steady profile.


\section{Advanced topic: Macroscopic reversibility by decomposing the equilibrium trajectory ensemble}
\label{app:revers}

Many of us are aware of the intrinsic time reversibility of Newtonian mechanics whereby any constant-energy trajectory $\xvec(t)$ can be ``played backwards'' to yield another physically valid trajectory.
There is an analogous condition on a stochastic trajectory, which can be derived from detailed balance. \cite{crooks1998nonequilibrium}
However, the conditions for reversibility under more realistic circumstances involving a \emph{distribution} of initial and final configurations require the trajectory ensemble picture. \cite{bhatt2011revers}

\begin{figure}
    \centering
    \includegraphics[width=7cm]{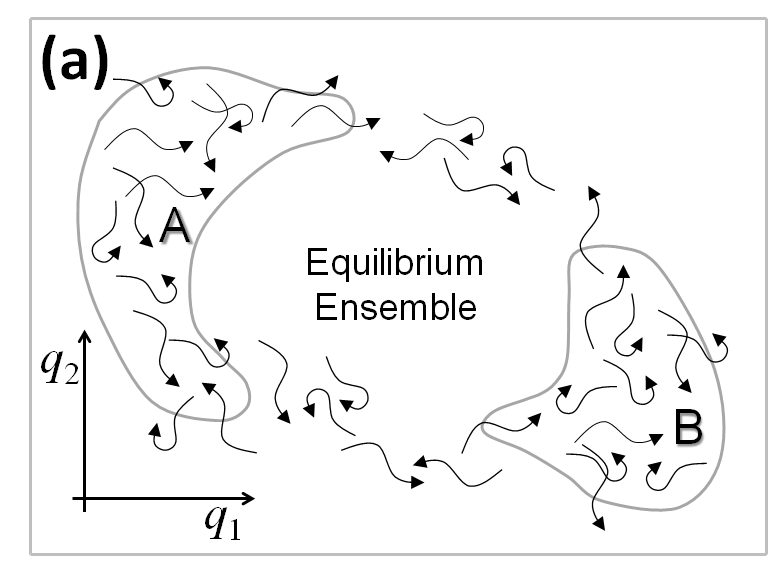}
    \includegraphics[width=7cm]{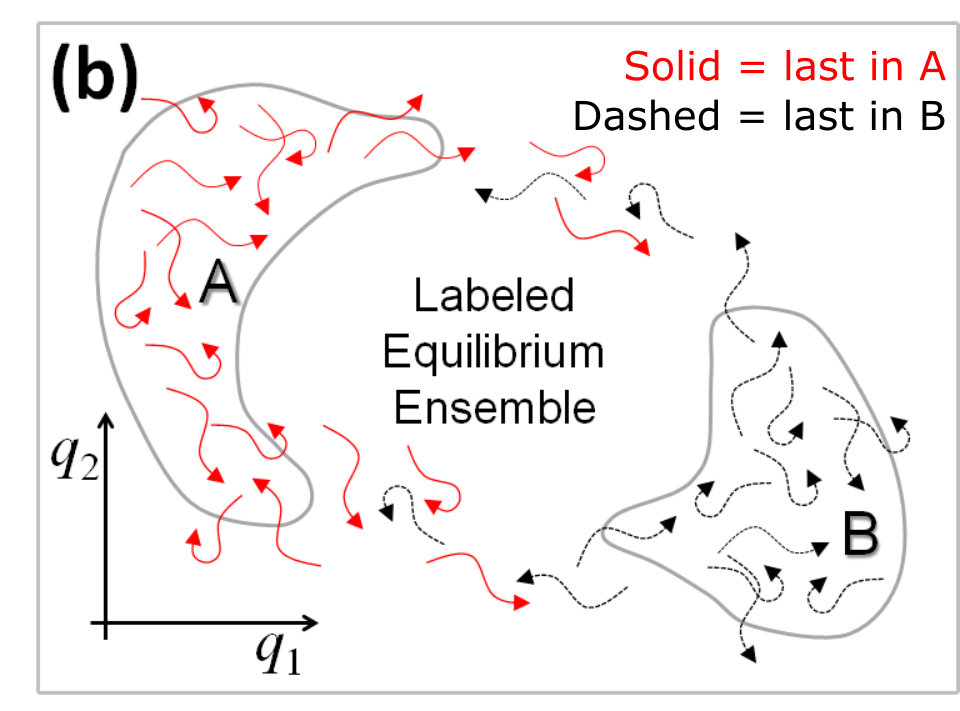}
    \caption{An exact decomposition of the equilibrium trajectory ensemble.
    (a) The equilibrium ensemble, consisting of a large number of \emph{independent} trajectories projected onto the schematic coordinates $q_1$ and $q_2$.  Transitions between macrostates A and B (gray outlines) occur via two pathways, upper and lower.
    (b) Decomposition of the equilibrium ensemble based on which of two macrostates, A and B, has been visited most recently.  These two directional ensembles, `last in A' or `A-to-B' (red, solid lines) and `last in B' or `B-to-A' (black, dashed lines), are non-equilibrium steady states.   Arrow tips represent the same time point for all trajectories and arrow tails represent the most recent history, but all history is assumed to be known.  
    Reprinted with permission from Bhatt and Zuckerman, J. Chem. Theory Comput. \textbf{7}(8), 2520–2527 (2011).
    Copyright 2011, American Chemical Society.
    \cite{bhatt2011revers}
    }
    \label{fig:equilrevers}
\end{figure}

We start by considering an equilibrium ensemble of trajectories: see Fig.\ \ref{fig:equilrevers}(a).
The equilibrium trajectory ensemble is defined by a set of completely independent systems/trajectories for times $t>t_0$ given that at $t_0$, the set of phase-space points $\xvec(t_0)$ is equilibrium-distributed -- i.e., according to the Boltzmann factor.
(We don't need to worry about how equilibrium was produced.)
If the phase points are equilibrium-distributed at time $t_0$, they will remain equilibrium-distributed thereafter.
This is because the Markovian stochastic dynamics that generates equilibrium also maintains it ... which is why we call it equilibrium in the first place. \cite{zuckerman2010statistical}
%

As sketched in Fig.\ \ref{fig:equilrevers}, the equilibrium ensemble at any time $t$ can be \emph{exactly} decomposed into two parts based on a history-labeling process.
\cite{van2003novel,bhatt2011revers}
Specifically, based on two arbitrary non-overlapping macrostates A and B, each trajectory can be assigned to the A-to-B set -- a.k.a ``last-in-A'' set -- if it currently occupies state A or was more recently in A than B, with the remaining trajectories in the B-to-A direction.
This construction requires ``omniscience,'' in the sense of knowing the full history of each trajectory, so it  is something of a thought experiment.
Note that each of these directional trajectory subsets is automatically maintained as a \emph{non-equilibrium} steady state: when an A-to-B trajectory enters B its label switches to B-to-A, but the overall equilibrium condition ensures that equal numbers of trajectories will switch labels per unit time. \cite{bhatt2011revers}

We're now in a position to understand reversibility, building on the defining process of equilibrium, detailed balance. \cite{zuckerman2010statistical}
As a reminder, detailed balance implies there is zero net flow between \emph{any} pair of ``microstates, '' i.e., small phase-space volumes.
In the context of the two uni-directional steady states (A-to-B and B-to-A), detailed balance gives us a tool to consider two non-overlapping mechanistic ``pathways'' --- arbitrary tubes of phase points connecting A and B --- e.g., upper and lower pathways in Fig.\ \ref{fig:equilrevers}.
If we place a (hyper-)surface transecting each tube, then there is a certain probability flowing per second through each surface in, say, the A-to-B steady state; call these $\sigma_1$ and $\sigma_2$.
By detailed balance, there is no net flow through either surface in equilibrium and so the flows in the B-to-A state must be equal and opposite.
Mechanistically, the ratio $\sigma_1/\sigma_2$ is the same in both directions: the fraction of events taking each pathway must be the same in both directions.
This is mechanistic reversibility.
Fuller details and illustrations can be found in earlier work. \cite{bhatt2011revers}

A key point is that the preceding discussion is strictly based on the detailed-balance property of equilibrium.
Thus, systems out of equilibrium should \emph{not} be expected to exhibit mechanistic reversibility.
This is true experimentally and theoretically.
Examples of systems not obeying reversibility would be if A and B states were prepared under different conditions (e.g., temperature, pH, ...) or, even under the same conditions, if the initial distribution in A or B did not mimic the process for constructing the directional steady states derived from equilibrium.
Specifically, in the A-to-B direction, trajectories should be initiated on the surface of A according to the distribution with which they would arrive from B in equilibrium, which is known as the ``EqSurf'' construction.
\cite{bhatt2011revers}
To put this informally, state A needs to be ``tricked'' into behaving as it would in equilibrium, so trajectories are started at the boundary of A as if they had arrived from B (i.e., were last in B) in equilibrium.

\end{document}